\documentclass[5p,times,twocolumn]{elsarticle}
\usepackage{lineno,hyperref}
\usepackage{subcaption}
\usepackage{tikz}
\usepackage{pgfplotstable}
\usepackage{pgfplots}
\usepackage{multirow}
\usepackage{float}
\usepackage{textcomp}
\usepackage{amsmath,amsfonts,amssymb,bbm}	
\usepackage{multicol}

\usepackage{siunitx}
\usepackage{graphicx} 
\usepackage{lscape} 
\usepackage{booktabs} 
\usepackage{footmisc} 
\usepackage{footnote} 
\makesavenoteenv{tabular} 
\makesavenoteenv{table} 

\usepackage[section]{placeins}

\modulolinenumbers[5]

\biboptions{authoryear}

\begin{document}

\begin{frontmatter}

\title{Fine-tuning deep learning model parameters for improved super-resolution of dynamic MRI with prior-knowledge}

\author[1,2,3]{Chompunuch Sarasaen\corref{equalcontribution}}
\author[1,3,4,5]{Soumick Chatterjee\corref{equalcontribution}}
\author[1,3]{Mario Breitkopf}
\author[2,3]{Georg Rose}
\author[4,5,7]{Andreas N{\"u}rnberger}
\author[1,3,6,7,8]{Oliver~Speck}

\cortext[equalcontribution]{C. Sarasaen and S. Chatterjee contributed equally to this work.}

\address[1]{Biomedical Magnetic Resonance, Otto von Guericke University Magdeburg, Germany}
\address[2]{Institute for Medical Engineering, Otto von Guericke University Magdeburg, Germany}
\address[3]{Research Campus STIMULATE, Otto von Guericke University Magdeburg, Germany}
\address[4]{Faculty of Computer Science, Otto von Guericke University Magdeburg, Germany}
\address[5]{Data and Knowledge Engineering Group, Otto von Guericke University Magdeburg, Germany}
\address[6]{German Center for Neurodegenerative Disease, Magdeburg, Germany}
\address[7]{Center for Behavioral Brain Sciences, Magdeburg, Germany}
\address[8]{Leibniz Institute for Neurobiology, Magdeburg, Germany}

\begin{abstract}
Dynamic imaging is a beneficial tool for interventions to assess physiological changes. Nonetheless during dynamic MRI, while achieving a high temporal resolution, the spatial resolution is compromised. To overcome this spatio-temporal trade-off, this research presents a super-resolution (SR) MRI reconstruction with prior knowledge based fine-tuning to maximise spatial information while reducing the required scan-time for dynamic MRIs. A U-Net based network with perceptual loss is trained on a benchmark dataset and fine-tuned using one subject-specific static high resolution MRI as prior knowledge to obtain high resolution dynamic images during the inference stage. 3D dynamic data for three subjects were acquired with different parameters to test the generalisation capabilities of the network. The method was tested for different levels of in-plane undersampling for dynamic MRI. The reconstructed dynamic SR results after fine-tuning showed higher similarity with the high resolution ground-truth, while quantitatively achieving statistically significant improvement. The average SSIM of the lowest resolution experimented during this research (6.25\% of the k-space) before and after fine-tuning were 0.939 $\pm$ 0.008 and 0.957 $\pm$ 0.006 respectively. This could theoretically result in an acceleration factor of 16, which can potentially be acquired in less than half a second. The proposed approach shows that the super-resolution MRI reconstruction with prior-information can alleviate the spatio-temporal trade-off in dynamic MRI, even for high acceleration factors.
\end{abstract}


\end{frontmatter}


\section{Introduction}
Magnetic Resonance Imaging (MRI) is in clinical use for a few decades with the advantage of non-ionising radiation, non-invasiveness and an excellent soft tissue contrast. Considering the clear visibility of tumours because of its high soft tissue contrast, together with real-time supervision (e.g. thermometry), MRI is a promising tool for interventions. The visualisation of lesions, as well as the needle paths, have to be acquired prior to any interventional procedures, so-called a planning scan or a preinterventional MR imaging~\citep{mahnken2009ct}. Furthermore, in MR-guided interventions, such as liver biopsy, it is necessary to continuously acquire data and reconstruct a series of images during interventions, in order to examine dynamic movements of internal organs~\citep{bernstein2004handbook}. A clear interpretable visualisation of the target lesion, surrounding tissues including risk structures is crucial during interventions. In order to achieve a high temporal resolution during dynamic MRIs, because of the inherently slow speed of image acquisition, the amount of data to be acquired has to be reduced - which may result in loss of spatial resolution. Although there are a number of techniques dealing with this spatio-temporal trade-off~\citep{tsao2003k,lustig2006kt,lustig2007sparse,jung2009k}, their speed of reconstruction creates a hindrance for real-time or near real-time imaging. Therefore, a compromise between spatial and temporal resolution is inevitable during real-time MRIs and needs to be mitigated.

The so-called super-resolution (SR) algorithms aim to restore images with high spatial resolution from the corresponding low resolution images. SR approaches have been widely used for various applications~\citep{zhang2014super,sajjadi2017enhancenet}, including for super-resolution of MRIs (SR-MRI)~\citep{van2012super,plenge2012super,isaac2015super}. Furthermore, deep learning based super-resolution reconstruction has been substantiated in recent times to be a successful tool for SR-MRI~\citep{zeng2018simultaneous,he2020super}, including for dynamic MRIs~\citep{qin2018convolutional,lyu2020cine}. However, most deep learning based methods need large training data sets and finding such training data -- matching with the data of the real-time acquisition that needs to be reconstructed in terms of contrast and sequence -- can be a challenging task. Using a training set significantly different than the test set can produce results of poor quality~\citep{wang2018deep,wilson2020survey}. Several techniques have been used to deal with the problem of small datasets in deep learning, such as data augmentation~\citep{perez2017effectiveness} and synthetic data generation~\citep{lateh2017handling,frid2018gan}. However, these methods rely on artificially modifying the data to increase the size of the dataset. Patch-based training can also help cope with the small dataset problem by splitting each data into smaller patches. This can effectively increase the number of samples in the dataset without artificially modifying the data~\citep{frid2017modeling}. The patch-based super-resolution (PBSR) techniques learn the mapping function from given corresponding pairs of high resolution and low resolution image patches~\citep{yang2014single}. 

This study proposes a PBSR reconstruction, aiming at addressing the problem of the lack of large abdominal datasets for training. This research intends to improve deep learning based super-resolution of dynamic MR images by incorporating prior images (planning scan). The network was trained on a publicly available abdominal dataset of 40 subjects, acquired using different sequences than the dynamic MR that is to be reconstructed. After that, the network was fine-tuned using a high resolution prior planning scan of the same subject as the dynamic acquisition. 


\subsection{Related works}
Super-resolution approaches have been employed for a wide variety of tasks, such as computer vision~\citep{shi2016real,dong2016accelerating,sajjadi2017enhancenet}, remote sensing~\citep{zhang2014super,ran2020remote}, face-related tasks~\citep{tappen2012bayesian,yu2018face} and medical applications~\citep{isaac2015super,huang2017simultaneousBrain}. Deep learning based methods have been widely used in recent times for performing super-resolution~\citep{dong2014learningSRCNN,zhu2014single,dong2016accelerating,ran2020remote}. Moreover, deep learning based techniques have been proven to be a successful tool for numerous applications in the field of MRI, including for performing MR reconstruction~\citep{wang2016accelerating,hyun2018deep,hammernik2018learning,chatterjeemri2019} and for SR-MRI~\citep{zeng2018simultaneous,liu2018fusing,chaudhari2018super,he2020super}. Different deep learning based SR-MRI ideas have been proposed for static brain MRI ~\citep{huang2017simultaneousBrain,tanno2017bayesianBrain,pham2017brain,zeng2018simultaneous,liu2018fusing,chen2018efficientBrain,deka2020sparseBrain,gu2020medsrgan}. Furthermore, deep learning based methods have additionally been shown to tackle the spatio-temporal trade-off~\citep{liang2020video}, also for dynamic cardiac MR reconstruction~\citep{qin2018convolutional,lyu2020cine}. 

Single-image super-resolution techniques are classified into the groups of prediction-based, edge-based, image statistical and patch-based methods~\citep{yang2014single}. PBSR can overcome the need of large training datasets as the actual training is done using patches, rather than on whole images. The PBSR methods have been applied to different tasks, including applications in medical imaging~\citep{manjon2010nonPatchSR,rousseau2010nonPatchSR,zhang2012reconstructionPatchSR,coupe2013collaborativePatchSR,jain2017patchSR}. By employing PBSR, the reconstruction procedure can be driven to cope with the limitation of available training abdominal MR data~\citep{tang2016pairwise,misra2020patch}. 

The U-Net~\citep{ronneberger2015u} model, which was originally proposed for image segmentation, over the past few years has been proven to solve various inverse problems as well~\citep{hyun2018deep,iqbal2019super,ghodrati2019mrPLoss}. ~\cite{iqbal2019super} developed a U-Net based architecture for SR reconstruction of MR spectroscopic images. ~\cite{hyun2018deep} reconstructed MRI utilising a 2D U-Net from zero-filled data, undersampled using uniform Cartesian sampling (GRAPPA-like) with a dense k-space centre. ~\cite{ghodrati2019mrPLoss} employed a U-Net model to test the performance of this network structure for cardiac MRI reconstruction. Due to the promising results shown in the papers mentioned earlier, the current paper proposed a 3D U-Net~\citep{cciccek20163d} based architecture for performing SR-MRI for abdominal dynamic images. 

Transfer learning is a technique for re-proposing or adapting a pre-trained model with fine-tuning~\citep{bengio2017deep}. With transfer learning, the network weights learned from one task can be used as pre-trained weights for another task, and then the network is trained (fine-tuned) for the new task. It has been widely used in data mining and machine learning ~\citep{dai2007transferring,li2009transfer,choi2017transfer,lee2018cleannetTransfer}. Transfer learning can address the issue of having insufficient training data~\citep{zhao2017research,kim2020effectiveness}. The fine-tuning process is known to improve the network's performance and can help to converge in less training epochs with smaller datasets~\citep{pan2009surveyTL}. One of the main research inquiries of applying transfer learning is "what to transfer". This current study thereby, utilises the specific knowledge of priors from a static planning image, which is usually acquired earlier to an interventional procedure. The incorporation of priors is meant to constrain anatomical structures in the fine-tuning process and to improve the data fidelity term in the regularisation process.

To determine the reconstruction error during training a deep learning model, between the model's prediction and the corresponding ground-truth images, the selection of a loss function is crucial. Pixel-based loss functions such as mean squared error (L2 loss) are commonly used for SR, however, in terms of perceptual quality, it often generates too smooth results which is caused due to the loss of high-frequency details~\citep{wang2003multiscale,wang2004imageSSIM,johnson2016perceptual,ledig2017photo}. Perceptual loss has shown potential to achieve high-quality images for image transformation tasks such as style transfer~\citep{johnson2016perceptual,gatys2016imagePlossStyleTransf}. For MRI reconstruction, \cite{ghodrati2019mrPLoss} have shown a comparative study of loss functions, such as perceptual loss (using VGG-16 as perceptual loss network), pixel-based loss (L1 and L2) and patch-wise structural dissimilarity (Dssim), for deep learning based cardiac MR reconstruction. They found that the results of the perceptual loss outperformed the other loss functions. Hence, in this work a combination of perceptual loss network with the mean absolute error (MAE) as the loss function was used, which is explained in section \ref{sec:ploss}.


\subsection{Problem statement}
Given a low resolution image $I_{LR}$ and a corresponding high resolution image $I_{HR}$, the reconstructed high resolution image $\hat{I}_{HR}$ can be recovered from a super-resolution reconstruction using the following equation~\citep{wang2020deep}:

\begin{equation}
    \hat{I}_{HR} =  \digamma(I_{LR}; \theta)
    \label{eqn:LR} 
\end{equation}

where $\digamma$ denotes the super-resolution model that maps the image counterparts and $\theta$ denotes the parameters of $\digamma$. The SR image reconstruction is an ill-posed problem, a network model can be trained to solve the objective function: 
{`I}

\begin{equation}
    \begin{split} \hat{\theta} = {arg\,min} 
    {\mathcal{L}(\hat{I}_{HR},I_{HR})+\lambda R(\theta)} \end{split}
    \label{eqn:hatx} 
\end{equation}

where the $\mathcal{L}(\hat{I}_{HR},I_{HR})$ denotes the loss function between the approximated HR image $\hat{I}_{HR}$ and ground-truth image $I_{HR}$, $R(\theta)$ is a regularisation term and $\lambda$ denotes the trade-off parameter.


\subsection{Contributions}
This paper presents a method to incorporate prior knowledge in deep-learning based super-resolution and its application in dynamic MRI. The main contributions are as follows:
\begin{itemize}
 \item This paper addresses the trade-off between the spatial and temporal resolution of dynamic MRI by incorporating a static high resolution scan as prior knowledge while performing spatial super-resolution on low-resolution images, effectively reducing the required scan-time per volume - which in turn will improve the temporal resolution.
 
\item A 3D U-Net model was first trained for the task of SR-MRI on a benchmark dataset and was fine-tuned using a subject-specific prior planning scan. 
  \item This paper further tackles the problem of the lack of high-resolution dynamic MRI for training in two ways:
  \begin{itemize}
    \item By using a static benchmark dataset for training, having different contrasts and resolutions than the target dynamic MRI; followed by fine-tuning using one static planning scan
    \item By patch-based super-resolution training and fine-tuning
  \end{itemize}
  \item To achieve realistic super-resolved images,  perceptual loss was used as the loss function for training and fine-tuning the model, which was calculated using a 3D perceptual loss network pre-trained on MR images.  
\end{itemize} 


\section{Methodology}
This paper proposes a framework of patch-based MR super-resolution based on U-Net, incorporating prior knowledge. The framework can be divided into three stages: main training, fine-tuning and inference. The U-Net model was initially trained with a benchmark dataset for main-training and then was fine-tuned using a subject-specific prior static scan. Finally in the inference stage, high resolution dynamic MRIs were reconstructed from low resolution scans. This chapter starts with the description of various datasets used in this research, then explains the network architecture followed by the implementation and training, finally explains the metrics used for evaluation.     


\subsection{Data}
\label{sec:data}
In this work, 3D abdominal MR volumes were artificially downsampled in-plane using MRUnder~\citep{soumick_chatterjee_2020_3901455}\footnote{MRUnder on Github: \url{https://github.com/soumickmj/MRUnder}} pipeline to simulate low resolution datasets. The low resolution data was generated by performing undersampling in-plane (phase-encoding and read-out direction) by taking the centre of k-space without zero-padding. The CHAOS challenge dataset~\citep{kavur2020chaos} (T1-Dual; in- and opposed-phase images) was used for the main training. During this phase, the dataset was split into training and validation set, with the ratio of 70:30. High resolution 3D static (breath-hold) and 3D "pseudo"-dynamic (free-breathing) scans for 10 time-points (TP) using T1w FLASH sequence were acquired for fine-tuning and inference respectively. Each time-point of the dynamic acquisition was treated as separate 3D Volume. Three healthy subjects were scanned with the same sequence but with different parameters on a 3T MRI (Siemens Magnetom Skyra). This aims to test the generalisation of the network. For each subject, the 3D static and the 3D dynamic scans were acquired in different sessions using the same sequences and parameters. The sequence parameters of the various datasets have been listed in Table \ref{MRIparams_CHAOS_3DDyn}. The CHAOS dataset (for main training), the 3D static scans (for fine-tuning) and the 3D dynamic scans (for inference) were artificially downsampled for three different levels of image resolution, by taking 25\%, 10\% and 6.25\% of the k-space centre, resulting in MR acceleration factor of 2, 3 and 4 respectively (considering undersampling only in the phase-encoding direction). This can be accelerated theoretically to a factor of 4, 9 and 16 respectively considering the amount of data used for the SR reconstruction. The effective resolutions and the roughly estimated acquisition times of the low resolution images were calculated from the corresponding high-resolution images, are reported in Table \ref{3DDyn_ResTimeUnder}. The acquisition times were calculated as: 
\begin{equation}
    Acq Time = PE_n\times TR\times S_m
    \label{eqn:AcqTime}
\end{equation}
where $PE_n$ is the number of phase-encoding lines - which equals to the matrix dimension in this direction, $TR$ is the repetition time - a parameter of an MRI pulse sequence which defines the time between two consecutive radio-frequency pulses, and $S_m$ is the number of slices. Phase/slice oversampling, phase/slice resolutions and GRAPPA factor were taken into account while calculating $PE_n$. The low resolution images served as the input to the network and were compared against the high resolution ground-truth images. 


\begin{table*}
\centering
\caption{MRI acquisition parameters CHAOS dataset and subject-wise 3D dynamic scans. Static scans were performed using the same subject-wise sequence parameters as the dynamic scans for one time-point (TP), acquired at a different session.}
\label{MRIparams_CHAOS_3DDyn} 
\resizebox{\textwidth}{!}{
\begin{tabular}{lllll}
\toprule
                            & \begin{tabular}[c]{@{}l@{}}CHAOS\\ (40 Subjects)\end{tabular} & Subject 1 & Subject 2 & Subject 3 \\ \toprule
Sequence &
  \begin{tabular}[c]{@{}l@{}}T1 Dual In-Phase \\ \& Opposed-Phase\end{tabular} &
  T1w Flash 3D &
  T1w Flash 3D &
  T1w Flash 3D \\ \midrule
Resolution &
  \begin{tabular}[c]{@{}l@{}}1.44 x 1.44 x 5 - \\ 2.03 x 2.03 x 8 $mm^3$\end{tabular} &
  1.09 x 1.09 x 4 $mm^3$ &
  1.09 x 1.09 x 4 $mm^3$ &
  1.36 x 1.36 x 4 $mm^3$ \\ \midrule
FOV x, y, z &
  \begin{tabular}[c]{@{}l@{}}315 x 315 x 240 - \\ 520 x 520 x 280 $mm^3$\end{tabular} &
  280 x 210 x 160 $mm^3$ &
  280 x 210 x 160 $mm^3$ &
  350 x 262 x 176 $mm^3$ \\ \midrule
Encoding matrix &
  \begin{tabular}[c]{@{}l@{}}256 x 256 x 26 - \\ 400 x 400 x 50\end{tabular} &
  256 x 192 x 40 &
  256 x 192 x 40 &
  256 x 192 x 44 \\ \midrule
Phase/Slice oversampling    &   -    & 10/0 \%   & 10/0 \%   & 10/0 \%   \\ \midrule
TR/TE &
  \begin{tabular}[c]{@{}l@{}}110.17 - 255.54 ms / \\ 4.60 - 4.64 ms (In-Phase)\\ 2.30 ms (Opposed-Phase)\end{tabular} &
  2.34/0.93 ms &
  2.34/0.93 ms &
  2.23/0.93 ms \\ \midrule
Flip angle                  & \ang{80}    & \ang{8}     & \ang{8}     & \ang{8}     \\ \midrule
Bandwidth                   &   -    & 975 Hz/Px & 975 Hz/Px & 975 Hz/Px \\ \midrule
GRAPPA factor               & None  & 2         & None      & None      \\ \midrule
Phase/Slice partial Fourier &   -    & Off/Off   & Off/Off   & Off/Off   \\ \midrule
Phase/Slice resolution      &   -    & 75/65 \%  & 75/65 \%  & 50/64 \%  \\ \midrule
Fat suppression             &   -    & None      & On        & On        \\ \midrule
Time per TP                 &   -    & 5.53 sec  & 11.76 sec  & 8.01 sec  \\ \toprule
\end{tabular}
}
\end{table*}

\begin{table*}
\centering
\caption{Effective resolutions and estimated acquisition times (per TP) of the dynamic and static datasets after performing different levels of artificial undersampling.}
\label{3DDyn_ResTimeUnder} 
\begin{tabular}{lcccccc}
\toprule
                                                                          & \multicolumn{2}{c}{Subject 1} & \multicolumn{2}{c}{Subject 2} & \multicolumn{2}{c}{Subject 3} \\ \cline{2-7} 
 &
  \begin{tabular}[c]{@{}c@{}}Resolution\\ ($mm^3$)\end{tabular} &
  \begin{tabular}[c]{@{}c@{}}Acq. Time\\ ($sec$)\end{tabular} &
  \begin{tabular}[c]{@{}c@{}}Resolution\\ ($mm^3$)\end{tabular} &
  \begin{tabular}[c]{@{}c@{}}Acq. Time\\ ($sec$)\end{tabular} &
  \begin{tabular}[c]{@{}c@{}}Resolution\\ ($mm^3$)\end{tabular} &
  \begin{tabular}[c]{@{}c@{}}Acq. Time\\ ($sec$)\end{tabular} \\ \toprule
High Resolution Ground-truth & 1.09 x 1.09 x 4     & 4.81    & 1.09 x 1.09 x 4     & 9.61    & 1.36 x 1.36 x 4     & 6.62    \\
25\% of k-space                                                           & 2.19 x 2.19 x 4     & 1.22    & 2.19 x 2.19 x 4     & 2.43    & 2.73 x 2.73 x 4     & 1.65    \\
10\% of k-space                                                           & 3.50 x 3.50 x 4     & 0.47    & 3.50 x 3.50 x 4     & 0.94    & 4.38 x 4.38 x 4     & 0.66    \\
6.25\% of k-space                                                         & 4.38 x 4.38 x 4     & 0.28    & 4.38 x 4.38 x 4     & 0.56    & 5.47 x 5.47 x 4     & 0.42    \\ 
\toprule
\end{tabular}
\end{table*}



\begin{figure*} 
\centering
\includegraphics[width=0.90\textwidth]{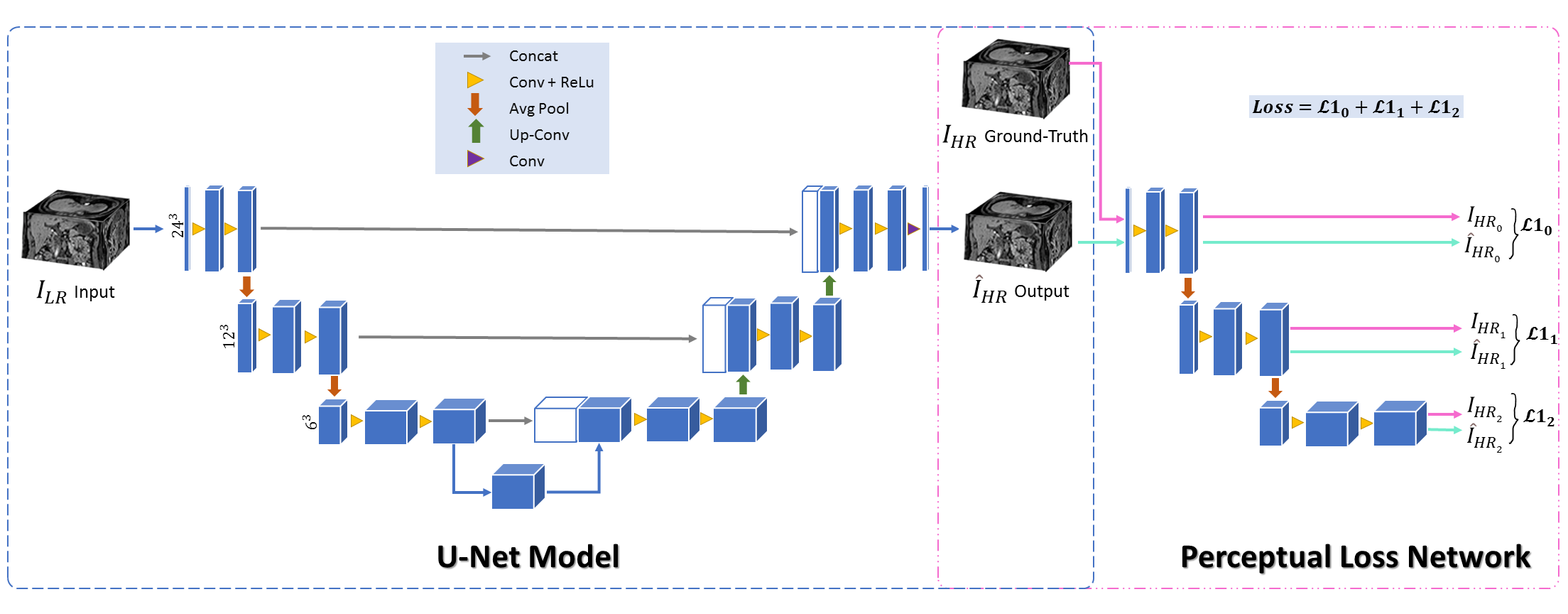}
\captionsetup{justification=centering}
\caption{The proposed network Architecture.}
\label{fig:ourArch}
\end{figure*}



\begin{figure*}
\centering
\includegraphics[width=0.8\textwidth]{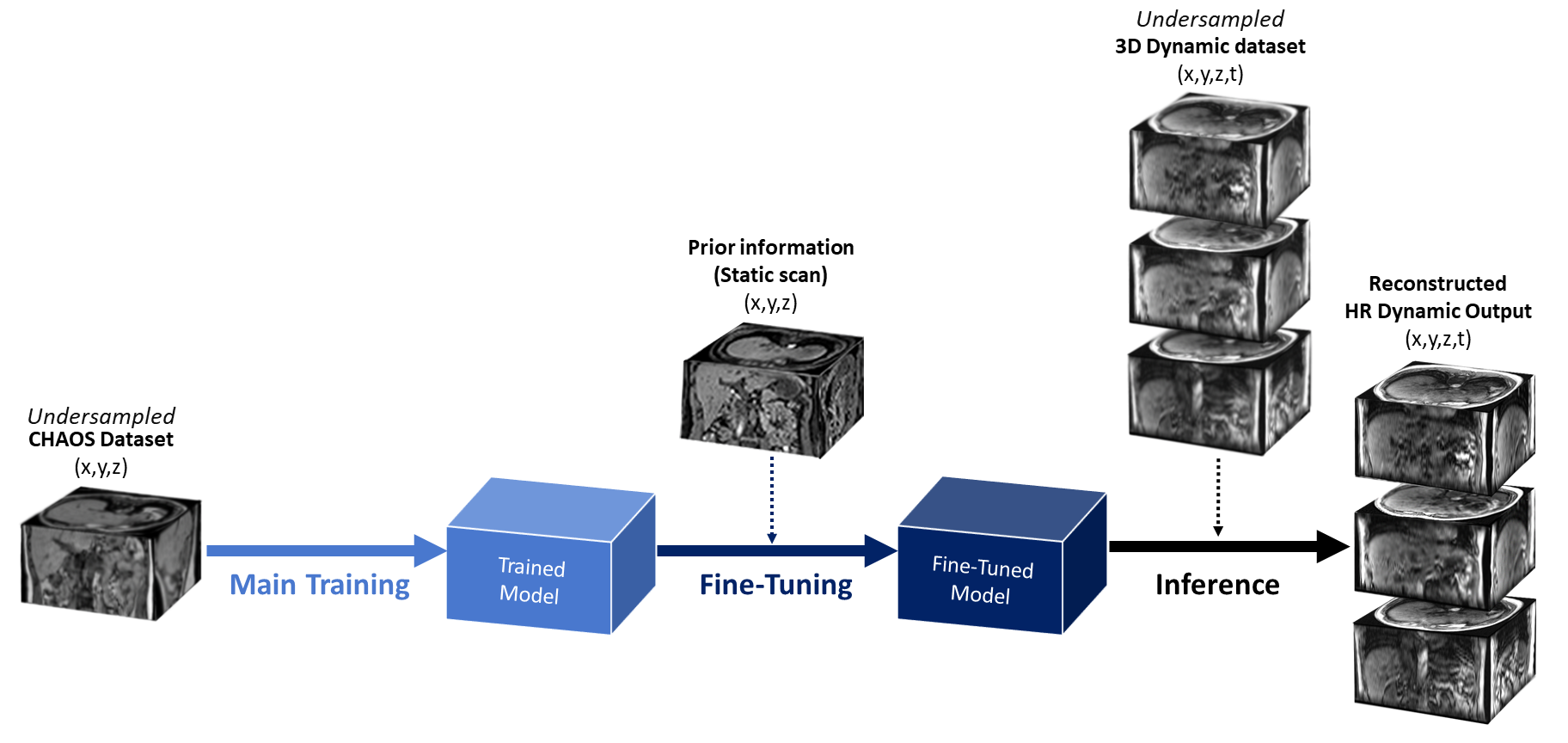}
\captionsetup{justification=centering}
\caption{Method Overview.}
\label{fig:Overview}
\end{figure*}



\subsection{Network Architecture}
Fig \ref{fig:ourArch} portrays the proposed network architecture. In this work, a 3D U-Net based model~\citep{ronneberger2015u,cciccek20163d,sarasaen2020esmrmb} with perceptual loss network~\citep{chatterjee2020ds6} was employed for super-resolution reconstruction. The U-Net architecture consists of two main paths; contracting (encoding) and expanding (decoding). The contracting path consists of three blocks, each comprises of two convolutional layers and a ReLU activation function. The expanding path also consists of three blocks, but convolutional transpose layers were used instead of the convolutional layers. The training was performed using 3D patches of the volumes, with a patch size of $24^3$. 

The U-Net model requires the same matrix size for input and output, therefore, the input low-resolution image patches were interpolated using trilinear interpolation before supplying them to the network. The interpolation factor was determined based on the undersampling factor - thus the low-resolution input is modified to the same matrix size as the target high-resolution ground-truth.

Patch-based training may result in patching artefacts during inference. To remove these artefacts, the inference was performed with a stride of one and the overlapped portions were averaged after reconstruction.

\subsection{Model Implementation and Network Training}
\label{sec:setup}
Fig \ref{fig:Overview} shows the method overview. The main training was performed using 3D patches of the 80 volumes (40 subjects, in-phase and opposed-phase for each subject), with a patch size of $24^3$ and a stride of 6 for the slice dimension and 12 for the other dimensions. After that, the network was fine-tuned using a single 3D static scan of the same subject from an earlier session, labelled as x,y,z,t in Fig \ref{fig:Overview}. This static scan has the same resolution, contrast and volume coverage as the high resolution dynamic scan. The static and the dynamic scans were not co-registered to keep it similar to the real-life scenario and to keep the speed of inference fast. Fine-tuning and evaluations were performed with a patch size of $24^3$ and a stride of one. The implementation was done using PyTorch~\citep{NEURIPS2019_9015} and was trained using Nvidia Tesla V100 GPUs. The loss was minimised using the Adam optimiser with a learning rate of 1e-4. The main training was performed for 200 epochs. The network was fine-tuned for only one epoch, using the planning scan with a lower learning rate (1e-6).

\subsubsection{Perceptual Loss}
\label{sec:ploss}
Loss during the training and fine-tuning of the network was calculated with the help of perceptual loss~\citep{johnson2016perceptual}. The first three levels of the contraction path of a pre-trained (on 7T MRA scans, for vessel segmentation) frozen U-Net MSS model~\citep{chatterjee2020ds6} were used as the perceptual loss network (PLN) to extract features from the final super-resolved output of the model and the ground-truth images (refer to Fig.\ref{fig:ourArch}). Typically VGG-16 trained on three-channel RGB non-medical images (ImageNet Dataset) is used as PLN, even while working with medical images~\citep{ghodrati2019mrPLoss} as the PLN doesn't have to be trained on a similar dataset. In this research, the pre-trained network was chosen because it was originally trained on single-channel medical images, but of different contrast and organ; and it was hypothesised that using a network trained as such will be more suitable than a network trained on three-channel images. The extracted features from the model's output and from the ground-truth images were compared using mean absolute error (L1 loss). The losses obtained at each level for each feature were then added together and backpropagated.


\subsection{Evaluation Metrics}
To evaluate the quality of the reconstructed images against the ground-truth HR images, two of the most commonly used metrics for evaluating image quality were selected, namely the structural similarity index (SSIM)\citep{wang2004imageSSIM} and the peak signal-to-noise ratio (PSNR). 

For perceptual quality assessment, the accuracy of the reconstructed images was compared to the ground truth using SSIM, which is based on the computations of luminance, contrast and structure terms between image x and y:

\begin{equation}
    SSIM (x,y) = \frac{(2\mu_x\mu_y+c_1)(2\sigma_{xy}+c_2)}{(\mu_x^2+\mu_y^2+c_1)(\sigma_x^2+\sigma_y^2+c_2)}
    \label{eqn:SSIM} 
\end{equation}
where $\mu_x, \mu_y, \sigma_x, \sigma_y$ and $\sigma_{xy}$ are the local means, standard deviations, and cross-covariance for images $x$ and $y$, respectively. $c_{1}=(k_{1}L)^{2}$ and $c_{2}=(k_{2}L)^{2}$, where $L$ is the dynamic range of the pixel-values, $k_{1}=0.01$ and $k_{2}=0.03$.

Additionally, the performance of the model was measured statistically with PSNR. It is calculated via the mean-square \\error~(MSE) as: 
\begin{equation}
    PSNR = 10 \log_{10} \left(\frac{R^2}{MSE}\right)
    \label{eqn:PSNR} 
\end{equation}
where $R$ is the maximum fluctuation in the input image.


\section{Results and Discussion}
Performance of the model was evaluated for three different levels of undersampling: by taking 25\%, 10\% and 6.25\% of the k-space centre. The network was tested before and after fine-tuning using 3D dynamic MRI. The proposed approach was compared against the low resolution input, the traditional trilinear interpolation, area interpolation (based on adaptive average pooling, as implemented in PyTorch\footnote{PyTorch Interpolation: \url{https://pytorch.org/docs/stable/generated/torch.nn.functional.interpolate.html}}) and finally against the most widely used technique in clinical MRIs - Fourier interpolation of the input (zero-padded k-space, also known as the sinc interpolation). 

\begin{figure*}
\centering
\includegraphics[width=\textwidth]{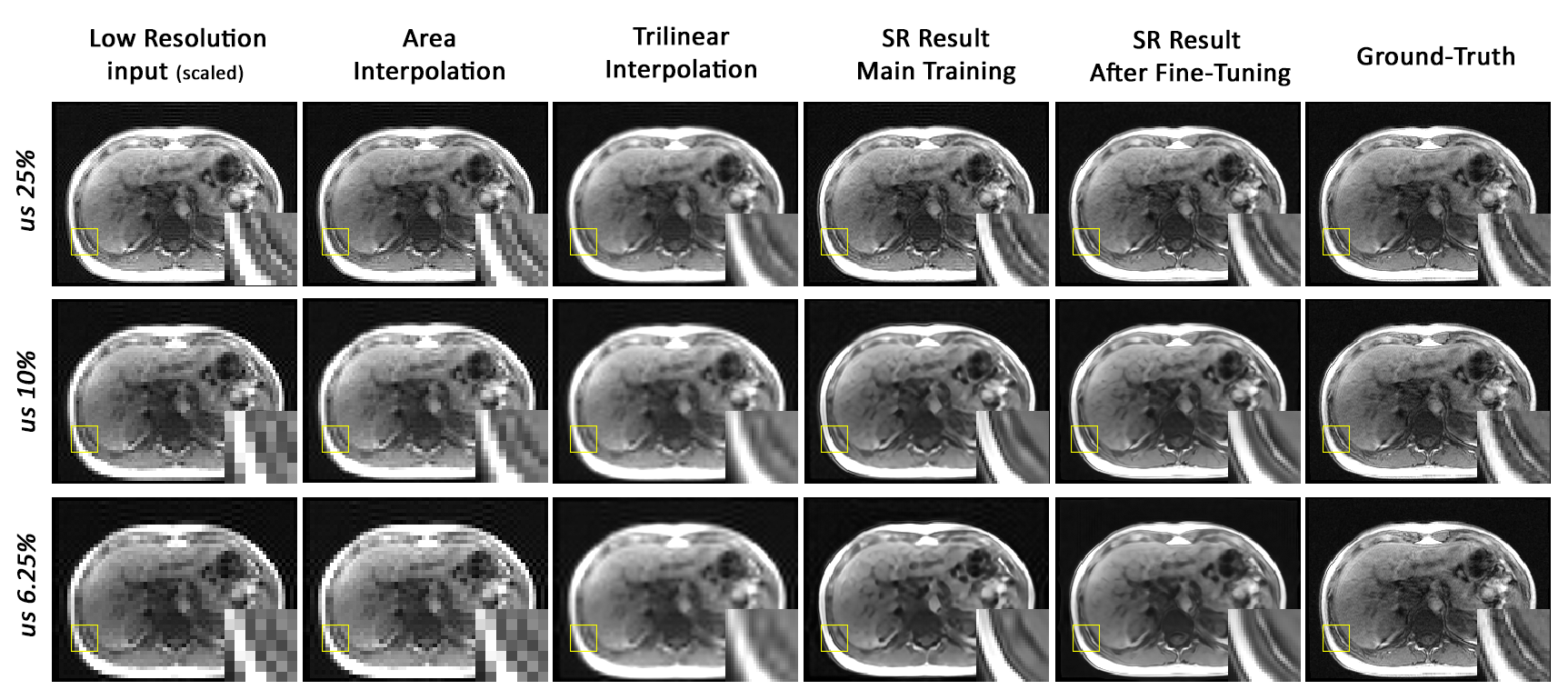}
\captionsetup{justification=centering}
\caption{Comparative results of low resolution (25\%, 10\% and 6.25\% of k-space) 3D Dynamic data of the same slice. From left to right: low resolution images (scaled-up), Interpolated input (Trilinear), super-resolution results of the main training (SR Results Main Training), super-resolution results after the fine-tune (SR After Fine-Tuning) and ground-truth images.}
\label{fig:3DDyn_us25us10us6p25compared}
\end{figure*}

\begin{figure*}
\centering
\includegraphics[width=0.85\textwidth]{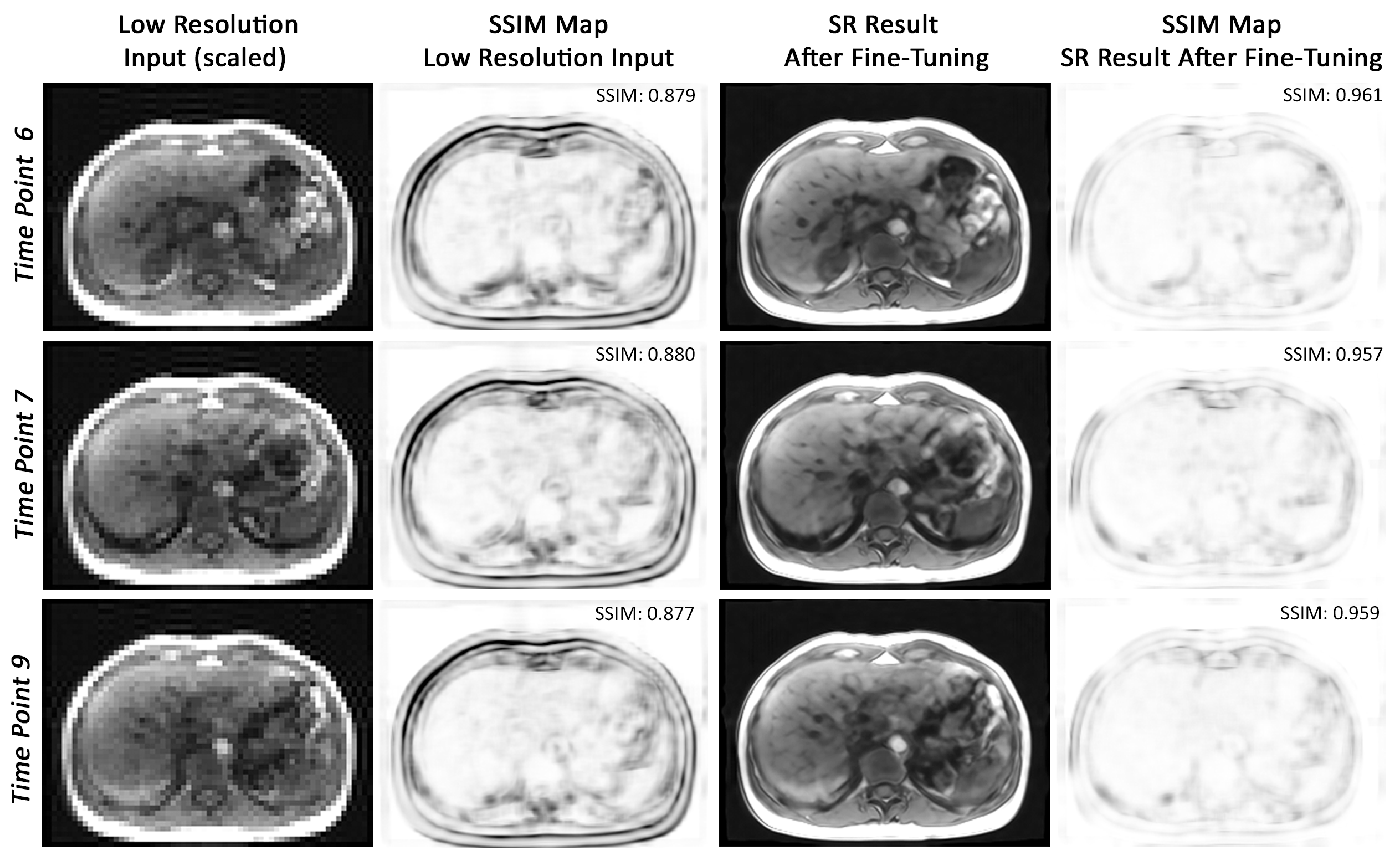}
\captionsetup{justification=centering}
\caption{An example comparison of the low resolution input of the 6.25\% of k-space with the super-resolution (SR) result after fine-tuning over three different time points, compared against the high resolution ground-truth using SSIM maps.}
\label{fig:input_output_us6p25_TP679_SSIM}
\end{figure*}

\begin{figure*}
\centering
\includegraphics[width=0.90\textwidth]{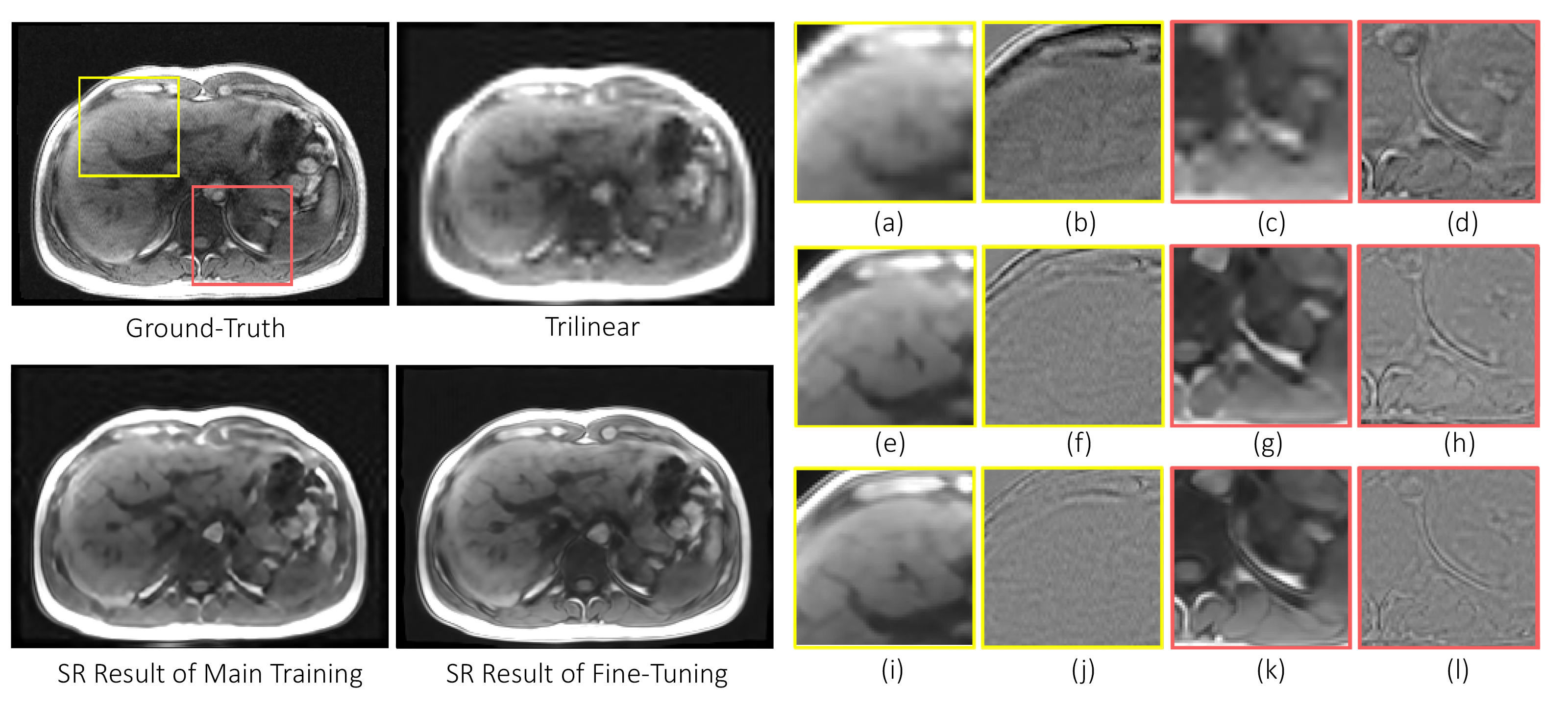}
\captionsetup{justification=centering}
\caption{An example from the reconstructed results, compared against its ground-truth (GT) for low resolution images from 6.25\% of k-space. From left to right, upper to lower: ground-truth, trilinear interpolation (input of the network), SR result of main training and SR result after fine-tuning. For the yellow ROI, (a-b): trilinear interpolation and the difference image from GT, (e-f): SR result of the main training and the difference image from GT and (i-j): SR result after fine-tuning and the difference image from GT. The images on the right part are identical examples for the red ROI.} 
\label{fig:roiUs6p25Subtract}
\end{figure*}

There was a noticeable improvement qualitatively and quantitatively while reconstructing low resolution data using the proposed method, even for only 6.25\% of the k-space. Fig \ref{fig:3DDyn_us25us10us6p25compared} shows the comparison qualitatively for the low resolution images by taking 25\%, 10\% and 6.25\% of the k-space. Fig \ref{fig:input_output_us6p25_TP679_SSIM} portrays the comparison of the low resolution input for 6.25\% of the k-space, the lowest resolution investigated during this study, with the SR result after fine-tuning over different time points. The SSIM maps were calculated against the high resolution ground-truth, which the respective SSIM value can be found on top of the image. Fig \ref{fig:roiUs6p25Subtract} illustrates the deviations of an example result from its corresponding ground-truth for two different regions of interest. It can be observed that SR results after fine-tuning could alleviate the undersampling artefacts, which are still present in the SR results of the main training, even for relatively low resolution images like 10\% and 6.25\% of the k-space. Consequently, the visibility of small details is improved.

\begin{figure*}
\centering
\includegraphics[width=\textwidth]{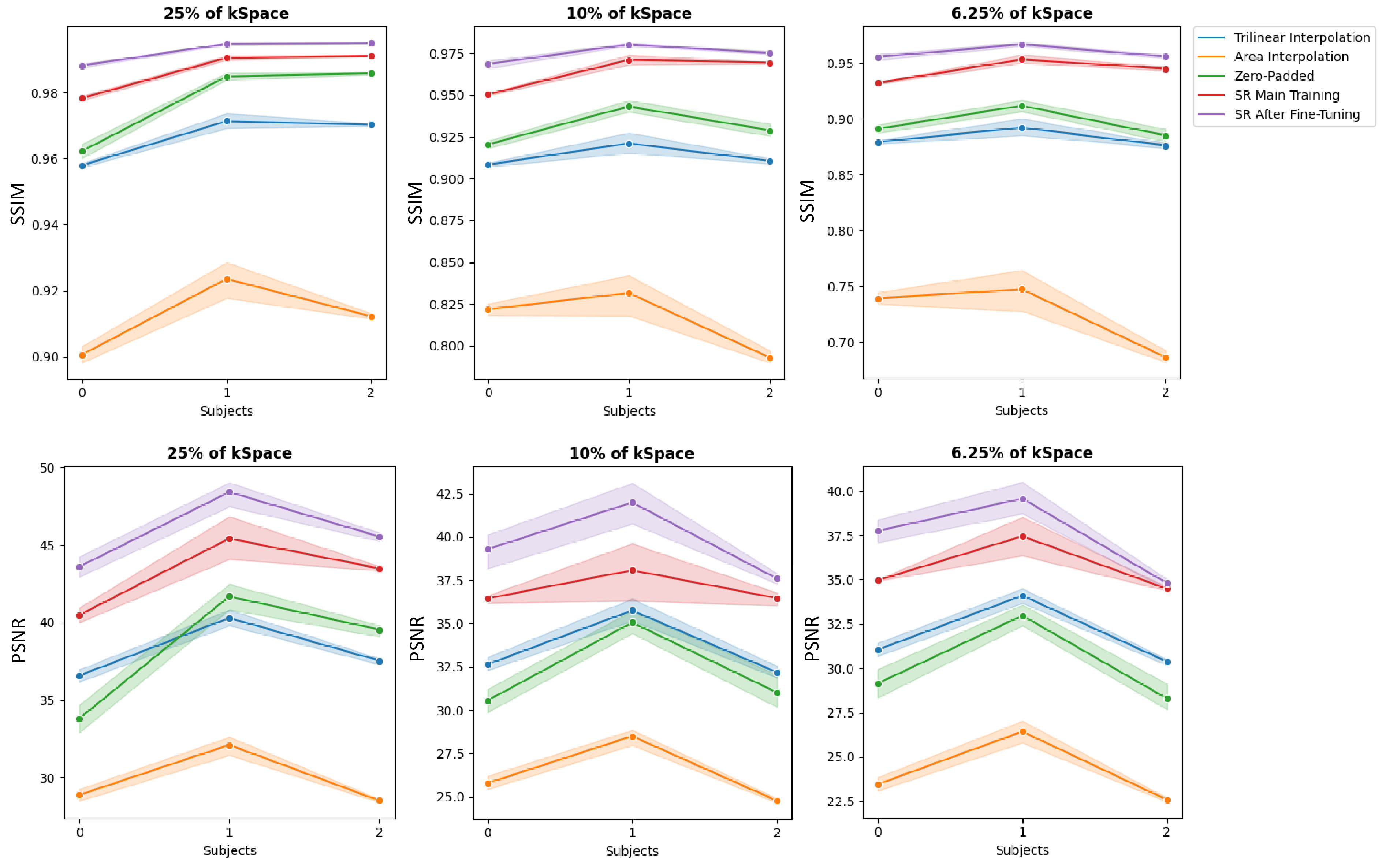}
\captionsetup{justification=centering}
\caption{Line plot showing the mean and 95\% confidence interval of the resultant SSIM and PSNR over the different time-points for each subject. The blue, orange, green, red, and violet lines represent the reconstruction results of trilinear interpolation, area interpolation, zero-padding (sinc interpolation), SR main training, and SR after fine-tuning, respectively. The upper row shows the SSIM values and the lower row shows the PSNR values.}
\label{fig:6graph_us25us10us6.25}
\end{figure*}

Additionally, for quantitative analysis, Table \ref{SSIMPSNRdiffSD_table_us25us10us6p25} displays the average and standard deviation (SD) of SSIM, PSNR and the SD of subtracted images for all time points for the dynamic datasets. Here, each time-point has been considered independent of each other as separate 3D volumes. Fig \ref{fig:6graph_us25us10us6.25} shows the distribution of the resultant metrics over all resolutions and subjects. The first row portrays the SSIM values and the second row shows the PSNR values. The columns denote 25\%, 10\%, and 6.25\% of the k-space, respectively. The blue, orange, green, red, and violet lines represent the reconstruction results of trilinear interpolation, area interpolation, zero-padding (sinc interpolation), SR main training, and SR after fine-tuning, whereas the thickness of the lines represents the standard deviation over time-points. It can be observed that the proposed method (SR after fine-tuning) significantly outperformed all the baseline methods experimented here.

Fine-tuning with the planning scan helped in obtaining sharper images and achieving a better edge delineation. Furthermore, the statistical significance of the improvements in terms of SSIM achieved by the model after fine-tuning was evaluated using paired t-test and Wilcoxon signed-rank test. Separate tests were performed considering all the different resolutions together and also by considering each resolution separately. It was observed that the improvement was statistically significant in every evaluated scenario (p-values were always less than 0.001). 

The acquisition time of high resolution 3D "pseudo"-dynamic reference data without parallel acquisition in this study was ten seconds and five seconds with GRAPPA factor two (Table \ref{3DDyn_ResTimeUnder}). These are not sufficient for real-time or near real-time applications and might lead to blurring in free-breathing subjects. This research shows the potential to acquire such volume with only minimal loss of spatial information in less than half a second.

The fine-tuning process took approximately eight hours to finish for each subject using the earlier mentioned setup (Section \ref{sec:setup}). Super-resolving each time-point took only a fraction of a second. The required time for fine-tuning and inference can further be reduced by reducing the patch-overlap (stride), though that might reduce the quality of the resultant super-resolved images. It can be further perceived that the network was able to produce results highly similar to the ground-truth (SSIM of 0.957) even while super-resolving 6.25\% of k-space, which can make the acquisition 16 times faster. Combining this fast acquisition speed with the inference speed of the method, this study can be extended to be used for real-time or near real-time MRIs during interventions. 

In the current study, only the centre of the k-space was used during undersampling, which results in loss of resolution without creating explicit image artefacts. Other undersampling patterns, such as variable density or GRAPPA-like uniform undersampling of higher spatial frequencies may be investigated in the future.


\begin{table*}
\centering
\caption{The average and the standard deviation of SSIM, PSNR, and SD of difference images with ground-truth. The table shows the results of different resolutions.}
\label{SSIMPSNRdiffSD_table_us25us10us6p25}
\resizebox{\textwidth}{!}{
\begin{tabular}{llllllllll}
\toprule
\multicolumn{1}{c}{\multirow{2}{*}{\textbf{Data}}} &
  \multicolumn{3}{c}{\textbf{25\% of k-space}} &
  \multicolumn{3}{c}{\textbf{10\% of k-space}} &
  \multicolumn{3}{c}{\textbf{6.25\% of k-space}} \\ \cline{2-10} 
\multicolumn{1}{c}{} &
  \multicolumn{1}{c}{\textbf{SSIM}} &
  \multicolumn{1}{c}{\textbf{PSNR}} &
  \textbf{diff SD} &
  \multicolumn{1}{c}{\textbf{SSIM}} &
  \multicolumn{1}{c}{\textbf{PSNR}} &
  \textbf{diff SD} &
  \multicolumn{1}{c}{\textbf{SSIM}} &
  \multicolumn{1}{c}{\textbf{PSNR}} &
  \multicolumn{1}{c}{\textbf{diff SD}} \\ \toprule
Area Interpolation &
  0.911$\pm$0.011 &
  29.721$\pm$1.948 &
  0.068$\pm$0.011 &
  0.814$\pm$0.018 &
  26.250$\pm$1.867 &
  0.071$\pm$0.013 &
  0.723$\pm$0.031 &
  24.092$\pm$1.964 &
  0.080$\pm$0.013 \\
Trilinear Interpolation &
  0.964$\pm$0.005 &
  37.680$\pm$1.770 &
  0.013$\pm$0.002 &
  0.906$\pm$0.007 &
  33.148$\pm$1.780 &
  0.022$\pm$0.004 &
  0.872$\pm$0.011 &
  31.504$\pm$1.786 &
  0.026$\pm$0.005 \\
Zero-padded &
  0.977$\pm$0.013 &
 37.980$\pm$4.078 &
  0.064$\pm$0.011 &
  0.926$\pm$0.009 &
  31.844$\pm$2.260 &
  0.067$\pm$0.013 &
  0.888$\pm$0.012 &
  29.803$\pm$2.147 &
  0.069$\pm$0.015 \\
SR Main Training &
  0.986$\pm$0.007 &
  42.781$\pm$2.424 &
  0.009$\pm$0.002 &
  0.961$\pm$0.009 &
  36.710$\pm$1.086 &
  0.014$\pm$0.002 &
  0.939$\pm$0.008 &
  35.377$\pm$1.653 &
  0.017$\pm$0.003 \\
SR Fine-tuning &
  \textit{\textbf{0.993$\pm$0.004}} &
  \textit{\textbf{45.706$\pm$2.169}} &
  \textit{\textbf{0.005$\pm$0.002}} &
  \textit{\textbf{0.973$\pm$0.005}} &
  \textit{\textbf{39.433$\pm$2.144}} &
  \textit{\textbf{0.007$\pm$0.001}} &
  \textit{\textbf{0.957$\pm$0.006}} &
  \textit{\textbf{37.306$\pm$2.357}} &
  \textit{\textbf{0.014$\pm$0.004}} \\ \toprule
\end{tabular}}
\end{table*}

It should be noted that the static planning scans and the actual dynamic scans during interventions are typically acquired with different sequences, with planning scans having higher contrast and resolution than the dynamic scans. This study was conducted using the same sequence for static and dynamic scans, but different resolutions and positions (different scan session). An additional experiment was performed by fine-tuning using a volumetric interpolated breath-hold examination (VIBE) sequence as planning scan for one subject. Super-resolving the dynamic low-resolution images from 6.25\% of the k-space resulted in 0.032 lower SSIM than using the identical sequence with higher resolution for fine-tuning. This may be a limitation of the current approach but requires further investigation.


\section{Conclusion and Future Work}
This research shows that fine-tuning with a subject-specific prior static scan can significantly improve the results of deep learning based super-resolution (SR) reconstruction. A 3D U-Net based model was trained with the help of perceptual loss to estimate the reconstruction error. The network model was initially trained using the CHAOS abdominal benchmark dataset and was then fine-tuned using a static high resolution prior scan. The model was used to obtain super-resolved high resolution 3D abdominal dynamic MRI from their corresponding low resolution images. Even though the network was trained using MRI sequences different than the reconstructed dynamic MRI, the SR results after fine-tuning showed higher similarity with the ground-truth images. The proposed method could overcome the spatio-temporal trade-off by improving the spatial resolution of the images without compromising the speed of acquisition. This approach could be applied to real-time dynamic acquisitions, such as for interventional MRIs, because of the high speed of inference of deep learning models. 

In the presented approach, a 3D U-Net was used as the network model, which needs interpolation as a pre-processing step. Therefore the reconstructed images could be suffering from interpolation errors. As future work, network models such as SRCNN, which do not need interpolation, will be studied. In addition, image resolutions lower than the already investigated ones will be studied to check the network's limitations. Moreover, clinical interventions are performed with devices, such as needle, which are not present in the planning scan. The authors plan to extend this research in future by evaluating on images with such devices.

\section*{Acknowledgements}
This work was conducted within the context of the International Graduate School MEMoRIAL at Otto von Guericke University (OVGU) Magdeburg, Germany, kindly supported by the European Structural and Investment Funds (ESF) under the programme "Sachsen-Anhalt WISSENSCHAFT Internationalisierung“ (project no. ZS/2016/08/80646).



\bibliography{mybibfile}

\begin{thebibliography}{72}
\expandafter\ifx\csname natexlab\endcsname\relax\def\natexlab#1{#1}\fi
\providecommand{\url}[1]{\texttt{#1}}
\providecommand{\href}[2]{#2}
\providecommand{\path}[1]{#1}
\providecommand{\DOIprefix}{doi:}
\providecommand{\ArXivprefix}{arXiv:}
\providecommand{\URLprefix}{URL: }
\providecommand{\Pubmedprefix}{pmid:}
\providecommand{\doi}[1]{\href{http://dx.doi.org/#1}{\path{#1}}}
\providecommand{\Pubmed}[1]{\href{pmid:#1}{\path{#1}}}
\providecommand{\bibinfo}[2]{#2}
\ifx\xfnm\relax \def\xfnm[#1]{\unskip,\space#1}\fi
\bibitem[{Bengio et~al.(2017)Bengio, Goodfellow and Courville}]{bengio2017deep}
\bibinfo{author}{Bengio, Y.}, \bibinfo{author}{Goodfellow, I.},
  \bibinfo{author}{Courville, A.}, \bibinfo{year}{2017}.
\newblock \bibinfo{title}{Deep learning}. volume~\bibinfo{volume}{1}.
\newblock \bibinfo{publisher}{MIT press Massachusetts, USA:}.
\bibitem[{Bernstein et~al.(2004)Bernstein, King and
  Zhou}]{bernstein2004handbook}
\bibinfo{author}{Bernstein, M.A.}, \bibinfo{author}{King, K.F.},
  \bibinfo{author}{Zhou, X.J.}, \bibinfo{year}{2004}.
\newblock \bibinfo{title}{Handbook of MRI pulse sequences}.
\newblock \bibinfo{publisher}{Elsevier}.
\bibitem[{Chatterjee(2020)}]{soumick_chatterjee_2020_3901455}
\bibinfo{author}{Chatterjee, S.}, \bibinfo{year}{2020}.
\newblock \bibinfo{title}{soumickmj/mrunder: Initial release}.
\newblock \DOIprefix\doi{10.5281/zenodo.3901455}.
\bibitem[{Chatterjee et~al.(2019)Chatterjee, Breitkopf, Sarasaen, Rose,
  Nürnberger and Speck}]{chatterjeemri2019}
\bibinfo{author}{Chatterjee, S.}, \bibinfo{author}{Breitkopf, M.},
  \bibinfo{author}{Sarasaen, C.}, \bibinfo{author}{Rose, G.},
  \bibinfo{author}{Nürnberger, A.}, \bibinfo{author}{Speck, O.},
  \bibinfo{year}{2019}.
\newblock \bibinfo{title}{A deep learning approach for reconstruction of
  undersampled cartesian and radial data}, in: \bibinfo{booktitle}{ESMRMB
  2019}.
\bibitem[{Chatterjee et~al.(2020)Chatterjee, Prabhu, Pattadkal, Bortsova,
  Sarasaen, Dubost, Mattern, de~Bruijne, Speck and
  N{\"u}rnberger}]{chatterjee2020ds6}
\bibinfo{author}{Chatterjee, S.}, \bibinfo{author}{Prabhu, K.},
  \bibinfo{author}{Pattadkal, M.}, \bibinfo{author}{Bortsova, G.},
  \bibinfo{author}{Sarasaen, C.}, \bibinfo{author}{Dubost, F.},
  \bibinfo{author}{Mattern, H.}, \bibinfo{author}{de~Bruijne, M.},
  \bibinfo{author}{Speck, O.}, \bibinfo{author}{N{\"u}rnberger, A.},
  \bibinfo{year}{2020}.
\newblock \bibinfo{title}{Ds6, deformation-aware semi-supervised learning:
  Application to small vessel segmentation with noisy training data}.
\newblock \bibinfo{journal}{arXiv preprint arXiv:2006.10802} .
\bibitem[{Chaudhari et~al.(2018)Chaudhari, Fang, Kogan, Wood, Stevens, Gibbons,
  Lee, Gold and Hargreaves}]{chaudhari2018super}
\bibinfo{author}{Chaudhari, A.S.}, \bibinfo{author}{Fang, Z.},
  \bibinfo{author}{Kogan, F.}, \bibinfo{author}{Wood, J.},
  \bibinfo{author}{Stevens, K.J.}, \bibinfo{author}{Gibbons, E.K.},
  \bibinfo{author}{Lee, J.H.}, \bibinfo{author}{Gold, G.E.},
  \bibinfo{author}{Hargreaves, B.A.}, \bibinfo{year}{2018}.
\newblock \bibinfo{title}{Super-resolution musculoskeletal mri using deep
  learning}.
\newblock \bibinfo{journal}{Magnetic resonance in medicine}
  \bibinfo{volume}{80}, \bibinfo{pages}{2139--2154}.
\bibitem[{Chen et~al.(2018)Chen, Shi, Christodoulou, Xie, Zhou and
  Li}]{chen2018efficientBrain}
\bibinfo{author}{Chen, Y.}, \bibinfo{author}{Shi, F.},
  \bibinfo{author}{Christodoulou, A.G.}, \bibinfo{author}{Xie, Y.},
  \bibinfo{author}{Zhou, Z.}, \bibinfo{author}{Li, D.}, \bibinfo{year}{2018}.
\newblock \bibinfo{title}{Efficient and accurate mri super-resolution using a
  generative adversarial network and 3d multi-level densely connected network},
  in: \bibinfo{booktitle}{International Conference on Medical Image Computing
  and Computer-Assisted Intervention}, \bibinfo{organization}{Springer}. pp.
  \bibinfo{pages}{91--99}.
\bibitem[{Choi et~al.(2017)Choi, Fazekas, Sandler and Cho}]{choi2017transfer}
\bibinfo{author}{Choi, K.}, \bibinfo{author}{Fazekas, G.},
  \bibinfo{author}{Sandler, M.}, \bibinfo{author}{Cho, K.},
  \bibinfo{year}{2017}.
\newblock \bibinfo{title}{Transfer learning for music classification and
  regression tasks}.
\newblock \bibinfo{journal}{arXiv preprint arXiv:1703.09179} .
\bibitem[{{\c{C}}i{\c{c}}ek et~al.(2016){\c{C}}i{\c{c}}ek, Abdulkadir,
  Lienkamp, Brox and Ronneberger}]{cciccek20163d}
\bibinfo{author}{{\c{C}}i{\c{c}}ek, {\"O}.}, \bibinfo{author}{Abdulkadir, A.},
  \bibinfo{author}{Lienkamp, S.S.}, \bibinfo{author}{Brox, T.},
  \bibinfo{author}{Ronneberger, O.}, \bibinfo{year}{2016}.
\newblock \bibinfo{title}{3d u-net: learning dense volumetric segmentation from
  sparse annotation}, in: \bibinfo{booktitle}{International conference on
  medical image computing and computer-assisted intervention},
  \bibinfo{organization}{Springer}. pp. \bibinfo{pages}{424--432}.
\bibitem[{Coup{\'e} et~al.(2013)Coup{\'e}, Manj{\'o}n, Chamberland, Descoteaux
  and Hiba}]{coupe2013collaborativePatchSR}
\bibinfo{author}{Coup{\'e}, P.}, \bibinfo{author}{Manj{\'o}n, J.V.},
  \bibinfo{author}{Chamberland, M.}, \bibinfo{author}{Descoteaux, M.},
  \bibinfo{author}{Hiba, B.}, \bibinfo{year}{2013}.
\newblock \bibinfo{title}{Collaborative patch-based super-resolution for
  diffusion-weighted images}.
\newblock \bibinfo{journal}{NeuroImage} \bibinfo{volume}{83},
  \bibinfo{pages}{245--261}.
\bibitem[{Dai et~al.(2007)Dai, Xue, Yang and Yu}]{dai2007transferring}
\bibinfo{author}{Dai, W.}, \bibinfo{author}{Xue, G.R.}, \bibinfo{author}{Yang,
  Q.}, \bibinfo{author}{Yu, Y.}, \bibinfo{year}{2007}.
\newblock \bibinfo{title}{Transferring naive bayes classifiers for text
  classification}, in: \bibinfo{booktitle}{AAAI}, pp.
  \bibinfo{pages}{540--545}.
\bibitem[{Deka et~al.(2020)Deka, Mullah, Datta, Lakshmi and
  Ganesan}]{deka2020sparseBrain}
\bibinfo{author}{Deka, B.}, \bibinfo{author}{Mullah, H.U.},
  \bibinfo{author}{Datta, S.}, \bibinfo{author}{Lakshmi, V.},
  \bibinfo{author}{Ganesan, R.}, \bibinfo{year}{2020}.
\newblock \bibinfo{title}{Sparse representation based super-resolution of mri
  images with non-local total variation regularization}.
\newblock \bibinfo{journal}{SN Computer Science} \bibinfo{volume}{1},
  \bibinfo{pages}{1--13}.
\bibitem[{Dong et~al.(2014)Dong, Loy, He and Tang}]{dong2014learningSRCNN}
\bibinfo{author}{Dong, C.}, \bibinfo{author}{Loy, C.C.}, \bibinfo{author}{He,
  K.}, \bibinfo{author}{Tang, X.}, \bibinfo{year}{2014}.
\newblock \bibinfo{title}{Learning a deep convolutional network for image
  super-resolution}, in: \bibinfo{booktitle}{European conference on computer
  vision}, \bibinfo{organization}{Springer}. pp. \bibinfo{pages}{184--199}.
\bibitem[{Dong et~al.(2016)Dong, Loy and Tang}]{dong2016accelerating}
\bibinfo{author}{Dong, C.}, \bibinfo{author}{Loy, C.C.}, \bibinfo{author}{Tang,
  X.}, \bibinfo{year}{2016}.
\newblock \bibinfo{title}{Accelerating the super-resolution convolutional
  neural network}, in: \bibinfo{booktitle}{European conference on computer
  vision}, \bibinfo{organization}{Springer}. pp. \bibinfo{pages}{391--407}.
\bibitem[{Frid-Adar et~al.(2017)Frid-Adar, Diamant, Klang, Amitai, Goldberger
  and Greenspan}]{frid2017modeling}
\bibinfo{author}{Frid-Adar, M.}, \bibinfo{author}{Diamant, I.},
  \bibinfo{author}{Klang, E.}, \bibinfo{author}{Amitai, M.},
  \bibinfo{author}{Goldberger, J.}, \bibinfo{author}{Greenspan, H.},
  \bibinfo{year}{2017}.
\newblock \bibinfo{title}{Modeling the intra-class variability for liver lesion
  detection using a multi-class patch-based cnn}, in:
  \bibinfo{booktitle}{International Workshop on Patch-based Techniques in
  Medical Imaging}, \bibinfo{organization}{Springer}. pp.
  \bibinfo{pages}{129--137}.
\bibitem[{Frid-Adar et~al.(2018)Frid-Adar, Diamant, Klang, Amitai, Goldberger
  and Greenspan}]{frid2018gan}
\bibinfo{author}{Frid-Adar, M.}, \bibinfo{author}{Diamant, I.},
  \bibinfo{author}{Klang, E.}, \bibinfo{author}{Amitai, M.},
  \bibinfo{author}{Goldberger, J.}, \bibinfo{author}{Greenspan, H.},
  \bibinfo{year}{2018}.
\newblock \bibinfo{title}{Gan-based synthetic medical image augmentation for
  increased cnn performance in liver lesion classification}.
\newblock \bibinfo{journal}{Neurocomputing} \bibinfo{volume}{321},
  \bibinfo{pages}{321--331}.
\bibitem[{Gatys et~al.(2016)Gatys, Ecker and
  Bethge}]{gatys2016imagePlossStyleTransf}
\bibinfo{author}{Gatys, L.A.}, \bibinfo{author}{Ecker, A.S.},
  \bibinfo{author}{Bethge, M.}, \bibinfo{year}{2016}.
\newblock \bibinfo{title}{Image style transfer using convolutional neural
  networks}, in: \bibinfo{booktitle}{Proceedings of the IEEE conference on
  computer vision and pattern recognition}, pp. \bibinfo{pages}{2414--2423}.
\bibitem[{Ghodrati et~al.(2019)Ghodrati, Shao, Bydder, Zhou, Yin, Nguyen, Yang
  and Hu}]{ghodrati2019mrPLoss}
\bibinfo{author}{Ghodrati, V.}, \bibinfo{author}{Shao, J.},
  \bibinfo{author}{Bydder, M.}, \bibinfo{author}{Zhou, Z.},
  \bibinfo{author}{Yin, W.}, \bibinfo{author}{Nguyen, K.L.},
  \bibinfo{author}{Yang, Y.}, \bibinfo{author}{Hu, P.}, \bibinfo{year}{2019}.
\newblock \bibinfo{title}{Mr image reconstruction using deep learning:
  evaluation of network structure and loss functions}.
\newblock \bibinfo{journal}{Quantitative imaging in medicine and surgery}
  \bibinfo{volume}{9}, \bibinfo{pages}{1516--1527}.
\bibitem[{Gu et~al.(2020)Gu, Zeng, Chen, Wei, Zhang, Chen, Li, Qin, Xie, Jiang
  et~al.}]{gu2020medsrgan}
\bibinfo{author}{Gu, Y.}, \bibinfo{author}{Zeng, Z.}, \bibinfo{author}{Chen,
  H.}, \bibinfo{author}{Wei, J.}, \bibinfo{author}{Zhang, Y.},
  \bibinfo{author}{Chen, B.}, \bibinfo{author}{Li, Y.}, \bibinfo{author}{Qin,
  Y.}, \bibinfo{author}{Xie, Q.}, \bibinfo{author}{Jiang, Z.}, et~al.,
  \bibinfo{year}{2020}.
\newblock \bibinfo{title}{Medsrgan: medical images super-resolution using
  generative adversarial networks}.
\newblock \bibinfo{journal}{Multimedia Tools and Applications}
  \bibinfo{volume}{79}, \bibinfo{pages}{21815--21840}.
\bibitem[{Hammernik et~al.(2018)Hammernik, Klatzer, Kobler, Recht, Sodickson,
  Pock and Knoll}]{hammernik2018learning}
\bibinfo{author}{Hammernik, K.}, \bibinfo{author}{Klatzer, T.},
  \bibinfo{author}{Kobler, E.}, \bibinfo{author}{Recht, M.P.},
  \bibinfo{author}{Sodickson, D.K.}, \bibinfo{author}{Pock, T.},
  \bibinfo{author}{Knoll, F.}, \bibinfo{year}{2018}.
\newblock \bibinfo{title}{Learning a variational network for reconstruction of
  accelerated mri data}.
\newblock \bibinfo{journal}{Magnetic resonance in medicine}
  \bibinfo{volume}{79}, \bibinfo{pages}{3055--3071}.
\bibitem[{He et~al.(2020)He, Lei, Fu, Mao, Curran, Liu and Yang}]{he2020super}
\bibinfo{author}{He, X.}, \bibinfo{author}{Lei, Y.}, \bibinfo{author}{Fu, Y.},
  \bibinfo{author}{Mao, H.}, \bibinfo{author}{Curran, W.J.},
  \bibinfo{author}{Liu, T.}, \bibinfo{author}{Yang, X.}, \bibinfo{year}{2020}.
\newblock \bibinfo{title}{Super-resolution magnetic resonance imaging
  reconstruction using deep attention networks}, in:
  \bibinfo{booktitle}{Medical Imaging 2020: Image Processing},
  \bibinfo{organization}{International Society for Optics and Photonics}. p.
  \bibinfo{pages}{113132J}.
\bibitem[{Huang et~al.(2017)Huang, Shao and
  Frangi}]{huang2017simultaneousBrain}
\bibinfo{author}{Huang, Y.}, \bibinfo{author}{Shao, L.},
  \bibinfo{author}{Frangi, A.F.}, \bibinfo{year}{2017}.
\newblock \bibinfo{title}{Simultaneous super-resolution and cross-modality
  synthesis of 3d medical images using weakly-supervised joint convolutional
  sparse coding}, in: \bibinfo{booktitle}{Proceedings of the IEEE Conference on
  Computer Vision and Pattern Recognition}, pp. \bibinfo{pages}{6070--6079}.
\bibitem[{Hyun et~al.(2018)Hyun, Kim, Lee, Lee and Seo}]{hyun2018deep}
\bibinfo{author}{Hyun, C.M.}, \bibinfo{author}{Kim, H.P.},
  \bibinfo{author}{Lee, S.M.}, \bibinfo{author}{Lee, S.}, \bibinfo{author}{Seo,
  J.K.}, \bibinfo{year}{2018}.
\newblock \bibinfo{title}{Deep learning for undersampled mri reconstruction}.
\newblock \bibinfo{journal}{Physics in Medicine \& Biology}
  \bibinfo{volume}{63}, \bibinfo{pages}{135007}.
\bibitem[{Iqbal et~al.(2019)Iqbal, Nguyen, Hangel, Motyka, Bogner and
  Jiang}]{iqbal2019super}
\bibinfo{author}{Iqbal, Z.}, \bibinfo{author}{Nguyen, D.},
  \bibinfo{author}{Hangel, G.}, \bibinfo{author}{Motyka, S.},
  \bibinfo{author}{Bogner, W.}, \bibinfo{author}{Jiang, S.},
  \bibinfo{year}{2019}.
\newblock \bibinfo{title}{Super-resolution 1h magnetic resonance spectroscopic
  imaging utilizing deep learning}.
\newblock \bibinfo{journal}{Frontiers in oncology} \bibinfo{volume}{9}.
\bibitem[{Isaac and Kulkarni(2015)}]{isaac2015super}
\bibinfo{author}{Isaac, J.S.}, \bibinfo{author}{Kulkarni, R.},
  \bibinfo{year}{2015}.
\newblock \bibinfo{title}{Super resolution techniques for medical image
  processing}, in: \bibinfo{booktitle}{2015 International Conference on
  Technologies for Sustainable Development (ICTSD)},
  \bibinfo{organization}{IEEE}. pp. \bibinfo{pages}{1--6}.
\bibitem[{Jain et~al.(2017)Jain, Sima, Sanaei~Nezhad, Hangel, Bogner, Williams,
  Van~Huffel, Maes and Smeets}]{jain2017patchSR}
\bibinfo{author}{Jain, S.}, \bibinfo{author}{Sima, D.M.},
  \bibinfo{author}{Sanaei~Nezhad, F.}, \bibinfo{author}{Hangel, G.},
  \bibinfo{author}{Bogner, W.}, \bibinfo{author}{Williams, S.},
  \bibinfo{author}{Van~Huffel, S.}, \bibinfo{author}{Maes, F.},
  \bibinfo{author}{Smeets, D.}, \bibinfo{year}{2017}.
\newblock \bibinfo{title}{Patch-based super-resolution of mr spectroscopic
  images: application to multiple sclerosis}.
\newblock \bibinfo{journal}{Frontiers in neuroscience} \bibinfo{volume}{11},
  \bibinfo{pages}{13}.
\bibitem[{Johnson et~al.(2016)Johnson, Alahi and
  Fei-Fei}]{johnson2016perceptual}
\bibinfo{author}{Johnson, J.}, \bibinfo{author}{Alahi, A.},
  \bibinfo{author}{Fei-Fei, L.}, \bibinfo{year}{2016}.
\newblock \bibinfo{title}{Perceptual losses for real-time style transfer and
  super-resolution}, in: \bibinfo{booktitle}{European conference on computer
  vision}, \bibinfo{organization}{Springer}. pp. \bibinfo{pages}{694--711}.
\bibitem[{Jung et~al.(2009)Jung, Sung, Nayak, Kim and Ye}]{jung2009k}
\bibinfo{author}{Jung, H.}, \bibinfo{author}{Sung, K.}, \bibinfo{author}{Nayak,
  K.S.}, \bibinfo{author}{Kim, E.Y.}, \bibinfo{author}{Ye, J.C.},
  \bibinfo{year}{2009}.
\newblock \bibinfo{title}{k-t focuss: a general compressed sensing framework
  for high resolution dynamic mri}.
\newblock \bibinfo{journal}{Magnetic Resonance in Medicine: An Official Journal
  of the International Society for Magnetic Resonance in Medicine}
  \bibinfo{volume}{61}, \bibinfo{pages}{103--116}.
\bibitem[{Kavur et~al.(2020)Kavur, Gezer, Bar{\i}{\c{s}}, Aslan, Conze, Groza,
  Pham, Chatterjee, Ernst, {\"O}zkan et~al.}]{kavur2020chaos}
\bibinfo{author}{Kavur, A.E.}, \bibinfo{author}{Gezer, N.S.},
  \bibinfo{author}{Bar{\i}{\c{s}}, M.}, \bibinfo{author}{Aslan, S.},
  \bibinfo{author}{Conze, P.H.}, \bibinfo{author}{Groza, V.},
  \bibinfo{author}{Pham, D.D.}, \bibinfo{author}{Chatterjee, S.},
  \bibinfo{author}{Ernst, P.}, \bibinfo{author}{{\"O}zkan, S.}, et~al.,
  \bibinfo{year}{2020}.
\newblock \bibinfo{title}{Chaos challenge-combined (ct-mr) healthy abdominal
  organ segmentation}.
\newblock \bibinfo{journal}{Medical Image Analysis} , \bibinfo{pages}{101950}.
\bibitem[{Kim et~al.(2020)Kim, Kim, Cho, Song, Lee, Ahn, Park, Gong and
  Kim}]{kim2020effectiveness}
\bibinfo{author}{Kim, Y.G.}, \bibinfo{author}{Kim, S.}, \bibinfo{author}{Cho,
  C.E.}, \bibinfo{author}{Song, I.H.}, \bibinfo{author}{Lee, H.J.},
  \bibinfo{author}{Ahn, S.}, \bibinfo{author}{Park, S.Y.},
  \bibinfo{author}{Gong, G.}, \bibinfo{author}{Kim, N.}, \bibinfo{year}{2020}.
\newblock \bibinfo{title}{Effectiveness of transfer learning for enhancing
  tumor classification with a convolutional neural network on frozen sections}.
\newblock \bibinfo{journal}{Scientific Reports} \bibinfo{volume}{10},
  \bibinfo{pages}{1--9}.
\bibitem[{Lateh et~al.(2017)Lateh, Muda, Yusof, Muda and
  Azmi}]{lateh2017handling}
\bibinfo{author}{Lateh, M.A.}, \bibinfo{author}{Muda, A.K.},
  \bibinfo{author}{Yusof, Z.I.M.}, \bibinfo{author}{Muda, N.A.},
  \bibinfo{author}{Azmi, M.S.}, \bibinfo{year}{2017}.
\newblock \bibinfo{title}{Handling a small dataset problem in prediction model
  by employ artificial data generation approach: A review}, in:
  \bibinfo{booktitle}{Journal of Physics: Conference Series}, p.
  \bibinfo{pages}{012016}.
\bibitem[{Ledig et~al.(2017)Ledig, Theis, Husz{\'a}r, Caballero, Cunningham,
  Acosta, Aitken, Tejani, Totz, Wang et~al.}]{ledig2017photo}
\bibinfo{author}{Ledig, C.}, \bibinfo{author}{Theis, L.},
  \bibinfo{author}{Husz{\'a}r, F.}, \bibinfo{author}{Caballero, J.},
  \bibinfo{author}{Cunningham, A.}, \bibinfo{author}{Acosta, A.},
  \bibinfo{author}{Aitken, A.}, \bibinfo{author}{Tejani, A.},
  \bibinfo{author}{Totz, J.}, \bibinfo{author}{Wang, Z.}, et~al.,
  \bibinfo{year}{2017}.
\newblock \bibinfo{title}{Photo-realistic single image super-resolution using a
  generative adversarial network}, in: \bibinfo{booktitle}{Proceedings of the
  IEEE conference on computer vision and pattern recognition}, pp.
  \bibinfo{pages}{4681--4690}.
\bibitem[{Lee et~al.(2018)Lee, He, Zhang and Yang}]{lee2018cleannetTransfer}
\bibinfo{author}{Lee, K.H.}, \bibinfo{author}{He, X.}, \bibinfo{author}{Zhang,
  L.}, \bibinfo{author}{Yang, L.}, \bibinfo{year}{2018}.
\newblock \bibinfo{title}{Cleannet: Transfer learning for scalable image
  classifier training with label noise}, in: \bibinfo{booktitle}{Proceedings of
  the IEEE Conference on Computer Vision and Pattern Recognition}, pp.
  \bibinfo{pages}{5447--5456}.
\bibitem[{Li et~al.(2009)Li, Yang and Xue}]{li2009transfer}
\bibinfo{author}{Li, B.}, \bibinfo{author}{Yang, Q.}, \bibinfo{author}{Xue,
  X.}, \bibinfo{year}{2009}.
\newblock \bibinfo{title}{Transfer learning for collaborative filtering via a
  rating-matrix generative model}, in: \bibinfo{booktitle}{Proceedings of the
  26th annual international conference on machine learning}, pp.
  \bibinfo{pages}{617--624}.
\bibitem[{Liang et~al.(2020)Liang, Du, Li, Xue, Wang, Kou and
  Wang}]{liang2020video}
\bibinfo{author}{Liang, M.}, \bibinfo{author}{Du, J.}, \bibinfo{author}{Li,
  L.}, \bibinfo{author}{Xue, Z.}, \bibinfo{author}{Wang, X.},
  \bibinfo{author}{Kou, F.}, \bibinfo{author}{Wang, X.}, \bibinfo{year}{2020}.
\newblock \bibinfo{title}{Video super-resolution reconstruction based on deep
  learning and spatio-temporal feature self-similarity}.
\newblock \bibinfo{journal}{IEEE Transactions on Knowledge and Data
  Engineering} .
\bibitem[{Liu et~al.(2018)Liu, Wu, Yu, Tang, Zhang and Zhou}]{liu2018fusing}
\bibinfo{author}{Liu, C.}, \bibinfo{author}{Wu, X.}, \bibinfo{author}{Yu, X.},
  \bibinfo{author}{Tang, Y.}, \bibinfo{author}{Zhang, J.},
  \bibinfo{author}{Zhou, J.}, \bibinfo{year}{2018}.
\newblock \bibinfo{title}{Fusing multi-scale information in convolution network
  for mr image super-resolution reconstruction}.
\newblock \bibinfo{journal}{Biomedical engineering online}
  \bibinfo{volume}{17}, \bibinfo{pages}{1--23}.
\bibitem[{Lustig et~al.(2007)Lustig, Donoho and Pauly}]{lustig2007sparse}
\bibinfo{author}{Lustig, M.}, \bibinfo{author}{Donoho, D.},
  \bibinfo{author}{Pauly, J.M.}, \bibinfo{year}{2007}.
\newblock \bibinfo{title}{Sparse mri: The application of compressed sensing for
  rapid mr imaging}.
\newblock \bibinfo{journal}{Magnetic Resonance in Medicine: An Official Journal
  of the International Society for Magnetic Resonance in Medicine}
  \bibinfo{volume}{58}, \bibinfo{pages}{1182--1195}.
\bibitem[{Lustig et~al.(2006)Lustig, Santos, Donoho and Pauly}]{lustig2006kt}
\bibinfo{author}{Lustig, M.}, \bibinfo{author}{Santos, J.M.},
  \bibinfo{author}{Donoho, D.L.}, \bibinfo{author}{Pauly, J.M.},
  \bibinfo{year}{2006}.
\newblock \bibinfo{title}{kt sparse: High frame rate dynamic mri exploiting
  spatio-temporal sparsity}, in: \bibinfo{booktitle}{Proceedings of the 13th
  annual meeting of ISMRM, Seattle}.
\bibitem[{Lyu et~al.(2020)Lyu, Shan, Xie, Li and Wang}]{lyu2020cine}
\bibinfo{author}{Lyu, Q.}, \bibinfo{author}{Shan, H.}, \bibinfo{author}{Xie,
  Y.}, \bibinfo{author}{Li, D.}, \bibinfo{author}{Wang, G.},
  \bibinfo{year}{2020}.
\newblock \bibinfo{title}{Cine cardiac mri motion artifact reduction using a
  recurrent neural network}.
\newblock \bibinfo{journal}{arXiv preprint arXiv:2006.12700} .
\bibitem[{Mahnken et~al.(2009)Mahnken, Ricke and Wilhelm}]{mahnken2009ct}
\bibinfo{author}{Mahnken, A.H.}, \bibinfo{author}{Ricke, J.},
  \bibinfo{author}{Wilhelm, K.E.}, \bibinfo{year}{2009}.
\newblock \bibinfo{title}{CT-and MR-guided Interventions in Radiology}.
  volume~\bibinfo{volume}{22}.
\newblock \bibinfo{publisher}{Springer}.
\bibitem[{Manj{\'o}n et~al.(2010)Manj{\'o}n, Coup{\'e}, Buades, Fonov, Collins
  and Robles}]{manjon2010nonPatchSR}
\bibinfo{author}{Manj{\'o}n, J.V.}, \bibinfo{author}{Coup{\'e}, P.},
  \bibinfo{author}{Buades, A.}, \bibinfo{author}{Fonov, V.},
  \bibinfo{author}{Collins, D.L.}, \bibinfo{author}{Robles, M.},
  \bibinfo{year}{2010}.
\newblock \bibinfo{title}{Non-local mri upsampling}.
\newblock \bibinfo{journal}{Medical image analysis} \bibinfo{volume}{14},
  \bibinfo{pages}{784--792}.
\bibitem[{Misra et~al.(2020)Misra, Crispim-Junior and Tougne}]{misra2020patch}
\bibinfo{author}{Misra, D.}, \bibinfo{author}{Crispim-Junior, C.},
  \bibinfo{author}{Tougne, L.}, \bibinfo{year}{2020}.
\newblock \bibinfo{title}{Patch-based cnn evaluation for bark classification},
  in: \bibinfo{booktitle}{European Conference on Computer Vision},
  \bibinfo{organization}{Springer}. pp. \bibinfo{pages}{197--212}.
\bibitem[{Pan and Yang(2009)}]{pan2009surveyTL}
\bibinfo{author}{Pan, S.J.}, \bibinfo{author}{Yang, Q.}, \bibinfo{year}{2009}.
\newblock \bibinfo{title}{A survey on transfer learning}.
\newblock \bibinfo{journal}{IEEE Transactions on knowledge and data
  engineering} \bibinfo{volume}{22}, \bibinfo{pages}{1345--1359}.
\bibitem[{Paszke et~al.(2019)Paszke, Gross, Massa, Lerer, Bradbury, Chanan,
  Killeen, Lin, Gimelshein, Antiga, Desmaison, Kopf, Yang, DeVito, Raison,
  Tejani, Chilamkurthy, Steiner, Fang, Bai and Chintala}]{NEURIPS2019_9015}
\bibinfo{author}{Paszke, A.}, \bibinfo{author}{Gross, S.},
  \bibinfo{author}{Massa, F.}, \bibinfo{author}{Lerer, A.},
  \bibinfo{author}{Bradbury, J.}, \bibinfo{author}{Chanan, G.},
  \bibinfo{author}{Killeen, T.}, \bibinfo{author}{Lin, Z.},
  \bibinfo{author}{Gimelshein, N.}, \bibinfo{author}{Antiga, L.},
  \bibinfo{author}{Desmaison, A.}, \bibinfo{author}{Kopf, A.},
  \bibinfo{author}{Yang, E.}, \bibinfo{author}{DeVito, Z.},
  \bibinfo{author}{Raison, M.}, \bibinfo{author}{Tejani, A.},
  \bibinfo{author}{Chilamkurthy, S.}, \bibinfo{author}{Steiner, B.},
  \bibinfo{author}{Fang, L.}, \bibinfo{author}{Bai, J.},
  \bibinfo{author}{Chintala, S.}, \bibinfo{year}{2019}.
\newblock \bibinfo{title}{Pytorch: An imperative style, high-performance deep
  learning library}, in: \bibinfo{editor}{Wallach, H.},
  \bibinfo{editor}{Larochelle, H.}, \bibinfo{editor}{Beygelzimer, A.},
  \bibinfo{editor}{d\textquotesingle Alch\'{e}-Buc, F.}, \bibinfo{editor}{Fox,
  E.}, \bibinfo{editor}{Garnett, R.} (Eds.), \bibinfo{booktitle}{Advances in
  Neural Information Processing Systems 32}. \bibinfo{publisher}{Curran
  Associates, Inc.}, pp. \bibinfo{pages}{8024--8035}.
\bibitem[{Perez and Wang(2017)}]{perez2017effectiveness}
\bibinfo{author}{Perez, L.}, \bibinfo{author}{Wang, J.}, \bibinfo{year}{2017}.
\newblock \bibinfo{title}{The effectiveness of data augmentation in image
  classification using deep learning}.
\newblock \bibinfo{journal}{arXiv preprint arXiv:1712.04621} .
\bibitem[{Pham et~al.(2017)Pham, Ducournau, Fablet and
  Rousseau}]{pham2017brain}
\bibinfo{author}{Pham, C.H.}, \bibinfo{author}{Ducournau, A.},
  \bibinfo{author}{Fablet, R.}, \bibinfo{author}{Rousseau, F.},
  \bibinfo{year}{2017}.
\newblock \bibinfo{title}{Brain mri super-resolution using deep 3d
  convolutional networks}, in: \bibinfo{booktitle}{2017 IEEE 14th International
  Symposium on Biomedical Imaging (ISBI 2017)}, \bibinfo{organization}{IEEE}.
  pp. \bibinfo{pages}{197--200}.
\bibitem[{Plenge et~al.(2012)Plenge, Poot, Bernsen, Kotek, Houston,
  Wielopolski, van~der Weerd, Niessen and Meijering}]{plenge2012super}
\bibinfo{author}{Plenge, E.}, \bibinfo{author}{Poot, D.H.},
  \bibinfo{author}{Bernsen, M.}, \bibinfo{author}{Kotek, G.},
  \bibinfo{author}{Houston, G.}, \bibinfo{author}{Wielopolski, P.},
  \bibinfo{author}{van~der Weerd, L.}, \bibinfo{author}{Niessen, W.J.},
  \bibinfo{author}{Meijering, E.}, \bibinfo{year}{2012}.
\newblock \bibinfo{title}{Super-resolution methods in mri: can they improve the
  trade-off between resolution, signal-to-noise ratio, and acquisition time?}
\newblock \bibinfo{journal}{Magnetic resonance in medicine}
  \bibinfo{volume}{68}, \bibinfo{pages}{1983--1993}.
\bibitem[{Qin et~al.(2018)Qin, Schlemper, Caballero, Price, Hajnal and
  Rueckert}]{qin2018convolutional}
\bibinfo{author}{Qin, C.}, \bibinfo{author}{Schlemper, J.},
  \bibinfo{author}{Caballero, J.}, \bibinfo{author}{Price, A.N.},
  \bibinfo{author}{Hajnal, J.V.}, \bibinfo{author}{Rueckert, D.},
  \bibinfo{year}{2018}.
\newblock \bibinfo{title}{Convolutional recurrent neural networks for dynamic
  mr image reconstruction}.
\newblock \bibinfo{journal}{IEEE transactions on medical imaging}
  \bibinfo{volume}{38}, \bibinfo{pages}{280--290}.
\bibitem[{Ran et~al.(2020)Ran, Xu, Zhao, Li and Du}]{ran2020remote}
\bibinfo{author}{Ran, Q.}, \bibinfo{author}{Xu, X.}, \bibinfo{author}{Zhao,
  S.}, \bibinfo{author}{Li, W.}, \bibinfo{author}{Du, Q.},
  \bibinfo{year}{2020}.
\newblock \bibinfo{title}{Remote sensing images super-resolution with deep
  convolution networks}.
\newblock \bibinfo{journal}{Multimedia Tools and Applications}
  \bibinfo{volume}{79}, \bibinfo{pages}{8985--9001}.
\bibitem[{Ronneberger et~al.(2015)Ronneberger, Fischer and
  Brox}]{ronneberger2015u}
\bibinfo{author}{Ronneberger, O.}, \bibinfo{author}{Fischer, P.},
  \bibinfo{author}{Brox, T.}, \bibinfo{year}{2015}.
\newblock \bibinfo{title}{U-net: Convolutional networks for biomedical image
  segmentation}, in: \bibinfo{booktitle}{International Conference on Medical
  image computing and computer-assisted intervention},
  \bibinfo{organization}{Springer}. pp. \bibinfo{pages}{234--241}.
\bibitem[{Rousseau et~al.(2010)Rousseau, Initiative
  et~al.}]{rousseau2010nonPatchSR}
\bibinfo{author}{Rousseau, F.}, \bibinfo{author}{Initiative, A.D.N.}, et~al.,
  \bibinfo{year}{2010}.
\newblock \bibinfo{title}{A non-local approach for image super-resolution using
  intermodality priors}.
\newblock \bibinfo{journal}{Medical image analysis} \bibinfo{volume}{14},
  \bibinfo{pages}{594--605}.
\bibitem[{Sajjadi et~al.(2017)Sajjadi, Scholkopf and
  Hirsch}]{sajjadi2017enhancenet}
\bibinfo{author}{Sajjadi, M.S.}, \bibinfo{author}{Scholkopf, B.},
  \bibinfo{author}{Hirsch, M.}, \bibinfo{year}{2017}.
\newblock \bibinfo{title}{Enhancenet: Single image super-resolution through
  automated texture synthesis}, in: \bibinfo{booktitle}{Proceedings of the IEEE
  International Conference on Computer Vision}, pp.
  \bibinfo{pages}{4491--4500}.
\bibitem[{Sarasaen et~al.(2020)Sarasaen, Chatterjee, Nürnberger and
  Speck}]{sarasaen2020esmrmb}
\bibinfo{author}{Sarasaen, C.}, \bibinfo{author}{Chatterjee, S.},
  \bibinfo{author}{Nürnberger, A.}, \bibinfo{author}{Speck, O.},
  \bibinfo{year}{2020}.
\newblock \bibinfo{title}{Super resolution of dynamic mri using deep learning,
  enhanced by prior-knowledge}, in: \bibinfo{booktitle}{37th Annual Scientific
  Meeting Congress of the European Society for Magnetic Resonance in Medicine
  and Biology, 33(Supplement 1): S03.04, S28-S29},
  \bibinfo{organization}{Springer}.
\newblock \DOIprefix\doi{10.1007/s10334-020-00874-0}.
\bibitem[{Shi et~al.(2016)Shi, Caballero, Husz{\'a}r, Totz, Aitken, Bishop,
  Rueckert and Wang}]{shi2016real}
\bibinfo{author}{Shi, W.}, \bibinfo{author}{Caballero, J.},
  \bibinfo{author}{Husz{\'a}r, F.}, \bibinfo{author}{Totz, J.},
  \bibinfo{author}{Aitken, A.P.}, \bibinfo{author}{Bishop, R.},
  \bibinfo{author}{Rueckert, D.}, \bibinfo{author}{Wang, Z.},
  \bibinfo{year}{2016}.
\newblock \bibinfo{title}{Real-time single image and video super-resolution
  using an efficient sub-pixel convolutional neural network}, in:
  \bibinfo{booktitle}{Proceedings of the IEEE conference on computer vision and
  pattern recognition}, pp. \bibinfo{pages}{1874--1883}.
\bibitem[{Tang and Shao(2016)}]{tang2016pairwise}
\bibinfo{author}{Tang, Y.}, \bibinfo{author}{Shao, L.}, \bibinfo{year}{2016}.
\newblock \bibinfo{title}{Pairwise operator learning for patch-based
  single-image super-resolution}.
\newblock \bibinfo{journal}{IEEE Transactions on Image Processing}
  \bibinfo{volume}{26}, \bibinfo{pages}{994--1003}.
\bibitem[{Tanno et~al.(2017)Tanno, Worrall, Ghosh, Kaden, Sotiropoulos,
  Criminisi and Alexander}]{tanno2017bayesianBrain}
\bibinfo{author}{Tanno, R.}, \bibinfo{author}{Worrall, D.E.},
  \bibinfo{author}{Ghosh, A.}, \bibinfo{author}{Kaden, E.},
  \bibinfo{author}{Sotiropoulos, S.N.}, \bibinfo{author}{Criminisi, A.},
  \bibinfo{author}{Alexander, D.C.}, \bibinfo{year}{2017}.
\newblock \bibinfo{title}{Bayesian image quality transfer with cnns: exploring
  uncertainty in dmri super-resolution}, in: \bibinfo{booktitle}{International
  Conference on Medical Image Computing and Computer-Assisted Intervention},
  \bibinfo{organization}{Springer}. pp. \bibinfo{pages}{611--619}.
\bibitem[{Tappen and Liu(2012)}]{tappen2012bayesian}
\bibinfo{author}{Tappen, M.F.}, \bibinfo{author}{Liu, C.},
  \bibinfo{year}{2012}.
\newblock \bibinfo{title}{A bayesian approach to alignment-based image
  hallucination}, in: \bibinfo{booktitle}{European conference on computer
  vision}, \bibinfo{organization}{Springer}. pp. \bibinfo{pages}{236--249}.
\bibitem[{Tsao et~al.(2003)Tsao, Boesiger and Pruessmann}]{tsao2003k}
\bibinfo{author}{Tsao, J.}, \bibinfo{author}{Boesiger, P.},
  \bibinfo{author}{Pruessmann, K.P.}, \bibinfo{year}{2003}.
\newblock \bibinfo{title}{k-t blast and k-t sense: dynamic mri with high frame
  rate exploiting spatiotemporal correlations}.
\newblock \bibinfo{journal}{Magnetic Resonance in Medicine: An Official Journal
  of the International Society for Magnetic Resonance in Medicine}
  \bibinfo{volume}{50}, \bibinfo{pages}{1031--1042}.
\bibitem[{Van~Reeth et~al.(2012)Van~Reeth, Tham, Tan and Poh}]{van2012super}
\bibinfo{author}{Van~Reeth, E.}, \bibinfo{author}{Tham, I.W.},
  \bibinfo{author}{Tan, C.H.}, \bibinfo{author}{Poh, C.L.},
  \bibinfo{year}{2012}.
\newblock \bibinfo{title}{Super-resolution in magnetic resonance imaging: a
  review}.
\newblock \bibinfo{journal}{Concepts in Magnetic Resonance Part A}
  \bibinfo{volume}{40}, \bibinfo{pages}{306--325}.
\bibitem[{Wang and Deng(2018)}]{wang2018deep}
\bibinfo{author}{Wang, M.}, \bibinfo{author}{Deng, W.}, \bibinfo{year}{2018}.
\newblock \bibinfo{title}{Deep visual domain adaptation: A survey}.
\newblock \bibinfo{journal}{Neurocomputing} \bibinfo{volume}{312},
  \bibinfo{pages}{135--153}.
\bibitem[{Wang et~al.(2016)Wang, Su, Ying, Peng, Zhu, Liang, Feng and
  Liang}]{wang2016accelerating}
\bibinfo{author}{Wang, S.}, \bibinfo{author}{Su, Z.}, \bibinfo{author}{Ying,
  L.}, \bibinfo{author}{Peng, X.}, \bibinfo{author}{Zhu, S.},
  \bibinfo{author}{Liang, F.}, \bibinfo{author}{Feng, D.},
  \bibinfo{author}{Liang, D.}, \bibinfo{year}{2016}.
\newblock \bibinfo{title}{Accelerating magnetic resonance imaging via deep
  learning}, in: \bibinfo{booktitle}{2016 IEEE 13th International Symposium on
  Biomedical Imaging (ISBI)}, \bibinfo{organization}{IEEE}. pp.
  \bibinfo{pages}{514--517}.
\bibitem[{Wang et~al.(2004)Wang, Bovik, Sheikh and
  Simoncelli}]{wang2004imageSSIM}
\bibinfo{author}{Wang, Z.}, \bibinfo{author}{Bovik, A.C.},
  \bibinfo{author}{Sheikh, H.R.}, \bibinfo{author}{Simoncelli, E.P.},
  \bibinfo{year}{2004}.
\newblock \bibinfo{title}{Image quality assessment: from error visibility to
  structural similarity}.
\newblock \bibinfo{journal}{IEEE transactions on image processing}
  \bibinfo{volume}{13}, \bibinfo{pages}{600--612}.
\bibitem[{Wang et~al.(2020)Wang, Chen and Hoi}]{wang2020deep}
\bibinfo{author}{Wang, Z.}, \bibinfo{author}{Chen, J.}, \bibinfo{author}{Hoi,
  S.C.}, \bibinfo{year}{2020}.
\newblock \bibinfo{title}{Deep learning for image super-resolution: A survey}.
\newblock \bibinfo{journal}{IEEE transactions on pattern analysis and machine
  intelligence} .
\bibitem[{Wang et~al.(2003)Wang, Simoncelli and Bovik}]{wang2003multiscale}
\bibinfo{author}{Wang, Z.}, \bibinfo{author}{Simoncelli, E.P.},
  \bibinfo{author}{Bovik, A.C.}, \bibinfo{year}{2003}.
\newblock \bibinfo{title}{Multiscale structural similarity for image quality
  assessment}, in: \bibinfo{booktitle}{The Thrity-Seventh Asilomar Conference
  on Signals, Systems \& Computers, 2003}, \bibinfo{organization}{Ieee}. pp.
  \bibinfo{pages}{1398--1402}.
\bibitem[{Wilson and Cook(2020)}]{wilson2020survey}
\bibinfo{author}{Wilson, G.}, \bibinfo{author}{Cook, D.J.},
  \bibinfo{year}{2020}.
\newblock \bibinfo{title}{A survey of unsupervised deep domain adaptation}.
\newblock \bibinfo{journal}{ACM Transactions on Intelligent Systems and
  Technology (TIST)} \bibinfo{volume}{11}, \bibinfo{pages}{1--46}.
\bibitem[{Yang et~al.(2014)Yang, Ma and Yang}]{yang2014single}
\bibinfo{author}{Yang, C.Y.}, \bibinfo{author}{Ma, C.}, \bibinfo{author}{Yang,
  M.H.}, \bibinfo{year}{2014}.
\newblock \bibinfo{title}{Single-image super-resolution: A benchmark}, in:
  \bibinfo{booktitle}{European Conference on Computer Vision},
  \bibinfo{organization}{Springer}. pp. \bibinfo{pages}{372--386}.
\bibitem[{Yu et~al.(2018)Yu, Fernando, Ghanem, Porikli and
  Hartley}]{yu2018face}
\bibinfo{author}{Yu, X.}, \bibinfo{author}{Fernando, B.},
  \bibinfo{author}{Ghanem, B.}, \bibinfo{author}{Porikli, F.},
  \bibinfo{author}{Hartley, R.}, \bibinfo{year}{2018}.
\newblock \bibinfo{title}{Face super-resolution guided by facial component
  heatmaps}, in: \bibinfo{booktitle}{Proceedings of the European Conference on
  Computer Vision (ECCV)}, pp. \bibinfo{pages}{217--233}.
\bibitem[{Zeng et~al.(2018)Zeng, Zheng, Cai, Yang, Zhang and
  Chen}]{zeng2018simultaneous}
\bibinfo{author}{Zeng, K.}, \bibinfo{author}{Zheng, H.}, \bibinfo{author}{Cai,
  C.}, \bibinfo{author}{Yang, Y.}, \bibinfo{author}{Zhang, K.},
  \bibinfo{author}{Chen, Z.}, \bibinfo{year}{2018}.
\newblock \bibinfo{title}{Simultaneous single-and multi-contrast
  super-resolution for brain mri images based on a convolutional neural
  network}.
\newblock \bibinfo{journal}{Computers in biology and medicine}
  \bibinfo{volume}{99}, \bibinfo{pages}{133--141}.
\bibitem[{Zhang et~al.(2014)Zhang, Yang, Zhang and Shen}]{zhang2014super}
\bibinfo{author}{Zhang, H.}, \bibinfo{author}{Yang, Z.},
  \bibinfo{author}{Zhang, L.}, \bibinfo{author}{Shen, H.},
  \bibinfo{year}{2014}.
\newblock \bibinfo{title}{Super-resolution reconstruction for multi-angle
  remote sensing images considering resolution differences}.
\newblock \bibinfo{journal}{Remote Sensing} \bibinfo{volume}{6},
  \bibinfo{pages}{637--657}.
\bibitem[{Zhang et~al.(2012)Zhang, Wu, Yap, Feng, Lian, Chen and
  Shen}]{zhang2012reconstructionPatchSR}
\bibinfo{author}{Zhang, Y.}, \bibinfo{author}{Wu, G.}, \bibinfo{author}{Yap,
  P.T.}, \bibinfo{author}{Feng, Q.}, \bibinfo{author}{Lian, J.},
  \bibinfo{author}{Chen, W.}, \bibinfo{author}{Shen, D.}, \bibinfo{year}{2012}.
\newblock \bibinfo{title}{Reconstruction of super-resolution lung 4d-ct using
  patch-based sparse representation}, in: \bibinfo{booktitle}{2012 IEEE
  Conference on Computer Vision and Pattern Recognition},
  \bibinfo{organization}{IEEE}. pp. \bibinfo{pages}{925--931}.
\bibitem[{Zhao(2017)}]{zhao2017research}
\bibinfo{author}{Zhao, W.}, \bibinfo{year}{2017}.
\newblock \bibinfo{title}{Research on the deep learning of the small sample
  data based on transfer learning}, in: \bibinfo{booktitle}{AIP Conference
  Proceedings}, \bibinfo{organization}{AIP Publishing LLC}. p.
  \bibinfo{pages}{020018}.
\bibitem[{Zhu et~al.(2014)Zhu, Zhang and Yuille}]{zhu2014single}
\bibinfo{author}{Zhu, Y.}, \bibinfo{author}{Zhang, Y.},
  \bibinfo{author}{Yuille, A.L.}, \bibinfo{year}{2014}.
\newblock \bibinfo{title}{Single image super-resolution using deformable
  patches}, in: \bibinfo{booktitle}{Proceedings of the IEEE Conference on
  Computer Vision and Pattern Recognition}, pp. \bibinfo{pages}{2917--2924}.

\end{thebibliography}

\end{document}